\documentclass[preprint]{aastex}

\begin{document}

\title{Perfecting the Photometric Calibration of the ACS CCD Cameras}
\author{Ralph~C.\ Bohlin\altaffilmark{1}} 
\altaffiltext{1}{Space Telescope Science Institute, 3700 San Martin Drive,
Baltimore, MD 21218, USA}
\email{bohlin@stsci.edu}

\begin{abstract}

Newly acquired data and improved data reduction algorithms mandate a fresh look
at the absolute flux calibration of the CCD cameras on the Hubble Space
Telescope (HST) Advanced Camera for Surveys (ACS). The goals are to achieve a
1\% accuracy and to make this calibration more accessible to the HST guest
investigator. Absolute fluxes from the
CALSPEC\footnote{http://www.stsci.edu/hst/observatory/crds/calspec.html}
database for three primary hot 30,000--60,000K WDs define the sensitivity
calibrations for the WFC and HRC filters. The external uncertainty for the
absolute flux is $\sim$1\%, while the internal consistency of the sensitivities
in the broadband ACS filters is $\sim$0.3\% among the three primary WD flux
standards. For stars as cool as K type, the agreement with the CALSPEC standards
is within 1\% at the WFC1-1K subarray position, which achieves the 1\% precision
goal for the first time. After making a small adjustment to the filter bandpass
for F814W, the 1\% precision goal is achieved over the full F814W WFC field of
view for stars of K type and hotter. New encircled energies and absolute
sensitivities replace the seminal results of Sirianni et al. that were published
in 2005. After implementing the throughput updates, synthetic predictions of the
WFC and HRC count rates for the average of the three primary WD standard stars
agree with the observations to 0.1\%.

\end{abstract}


\section{Introduction}

Photometric calibration uncertainties are the dominant source of error in
current type Ia supernova dark energy studies and other cosmology efforts; and
the precision of the constraints on the constants in the Einstein equations of
general relativity are directly related to the precision of absolute flux
calibrations \citep{scolnic14}.
 
The ACS CCD (charge coupled device) flux calibration was updated fairly recently
in a series of four Instrument Science
Reports\footnote{http://www.stsci.edu/hst/acs/documents/isrs} that covered the
philosophy and techniques, including charge transfer efficiency (CTE)
corrections, encircled energy (EE) corrections, changes in sensitivity with
time, and the absolute flux calibration itself
\citep{bohlinjay2011,bohlin2011,bohlinetal11, bohlin2012}. However, additional
data, updates to the processing, and revised fluxes for the primary standard
stars, which change by up to $\sim$1\% \citep{bohlin15}, mandate a fresh
analysis of the complete body of calibration observations. ACS has two CCD
channels and a Solar Blind Channel (SBC). The CCD channels are a Wide Field
Channel (WFC) and a High Resolution Channel (HRC), both of which ceased
functioning in 2007. However, the WFC was revived during Servicing Mission four
(SM4) in 2009.

Changes to the data
reduction with generally sub-percent effects include: updated bias reference
files, a better removal of the weak striping in data obtained with the new
electronics installed during SM4, a refined
correction for the sensitivity degradation with time, and a fix of a minor flaw
in the calculation of the sky background in the IDL $apphot.pro$ routine used to
extract the photometry of the standard stars from the \textit{\_crj.fits} files.
The low level striping in the post-SM4 WFC data is removed  using the median for
each row, which works nicely for the sparse standard star fields in the
1024x1024 subarrays after masking a 5.5x5.5\arcsec\ box centered on the star.
The Space Telecope Science Institute (STScI) pipeline data processing begins
with the \textit{\_raw.fits} data files, and then applies the flat field
correction along with other corrections associated with CCDs to make
\textit{\_flt.fits} files. If there are multiple exposures at the same pointing
that are taken to permit the rejection of cosmic ray hits, then the
\textit{\_crj.fits} files are also produced from the \textit{\_flt.fits} files.
If multiple exposures are taken of the same target but with small dithered
pointing offsets, then the \textit{\_flt.fits} data are combined with the
drizzle software task to make \textit{\_drz.fits} files that have both hot
pixels and cosmic ray rejection. The flat fields are designed to produce the
same response in each pixel when  observing a perfectly uniform diffuse source.
However, optical distortions cause variations in the size of the pixels in
arcsec on the sky; so that a multiplication by a pixel-area map (PAM) file is
required for the \textit{\_flt.fits} and \textit{\_crj.fits} files to correct
for the pixel size variation and make a uniform response anywhere in the field
for the same stellar point source. The PAM correction is built into the drizzle
software; and no PAM correction is needed for the \textit{\_drz.fits} files.

The logical flow of the derivation of the ACS flux calibration starts with
extracting the latest version of the standard pipeline data products from the
Mikulski Archive for Space Telescopes
(MAST)\footnote{http://archive.stsci.edu/hst/search.php}. Then, the aperture
photometry is extracted from the \textit{\_crj.fits} files for various aperture
radii. These photometry data files form the basis for subsequent analyses,
including derivation of the fractional encircled energy (EE) as a function of
radius, of the change in sensitivity as a function of time on orbit, and of the
absolute flux calibration. The main goal of this work is to establish this flux
calibration with better than a 1\% precision at the WFC1-1K reference point for
the heavily used broadband filters. The HST method of parameterizing the
instrumental calibration of flux sensitivities relies on pre-launch estimates
for the fractional thoughput or quantum efficiency of each element in the
optical path from the primary mirror through to the detector that registers the
photon events. Instrumental sensitivities depend directly on the product of all
these component efficiencies and the collecting area of the 2.4 m primary
mirror. In-flight updates to the pre-launch estimates are derived from
observations of standard stars with known flux distributions and  are
implemented via changes to the detector QE as a smooth function of wavelength or
by scaler correction factors to the filter throughput functions. In this work,
wavelength dependent corrections for two filter bandpass functions are also
derived. No new updates to the QE vs. wavelength are needed, so that the
calibration reference files that must be delivered for the pipeline data
reduction includes only files to define the changing sensitivity vs. time and
files that update the filter throughputs.

Section 2 reviews the observational data and its pipeline reduction, while
Section 3 covers the details of the bias subtraction. Section 4 augments the EE
fractions, which now include apertures of radii from 0.05--2\arcsec, i.e. 1--40
WFC pixels. Section 5 reviews the sensitivity change with time; Section 6
justifies a change in the F814W bandpass throughput function; and Section 7
covers the  absolute flux calibration for photometry. An analysis of the
problematic \textit{\_drz.fits} files appears in the Appendix and shows that
more than a quarter of the 334 \textit{\_drz} images of standard stars are
erroneous.

The original comprehensive ACS calibration paper \citep{sirianni05} also covered
these topics, except for changes in sensitivity with time and wavelength
dependent bandpass adjustments. While the \citep{sirianni05} EE results are for
the center of the 1024x1024 pixel HRC channel, those WFC EE values are averages
of data at the centers of the ACS CCD 1 and CCD 2, which are butted to make the 
4096x4096 pixel WFC detector. Any location can be used for the absolute flux
calibration,as long as the flat field is perfect in making all locations
equivalent. For efficient use of the ACS calibration time, the center of the
WFC1-1K subarray is the chosen flux calibration reference point for the WFC. The
flat fields are derived from dithered observations of 47 Tuc \citep{mack02};
and the pioneering pilot check of these flats with dithered observations of
bright standard stars with a
wide range of stellar color temperature is \citet{bohgrog} for F814W and F435W.
With the bandpass adjustment of Section 6, F814W is now uniform to 1\% for stars
of type K and hotter. For F435W, further analysis of these new flat field checks
will be forthcoming.

\section{Observations}

\begin{deluxetable}{lllrrr}   
\tablewidth{0pt}
\tablecolumns{6}
\tablecaption{Stellar Flux Standards}
\tablehead{
\multicolumn{3}{c}{~} & \multicolumn{3}{c}{Number of Visits} \\
\colhead{Star} &\colhead{Sp.Type\tablenotemark{a}} &\colhead{V}
&\colhead{WFC} &\colhead{HRC} &\colhead{STIS}}
\startdata
G191B2B     &DA.8    & 11.781 &8 &2-3 &12-17 \\
GD153       &DA1.2   & 13.346 &8 &2  &14-32 \\
GD71        &DA1     & 13.032 &7-8 &2  &17-34 \\
GRW+70$^{\circ}$5824 &DA3     & 12.77  &0 &4  &1-85\tablenotemark{b}  \\
BD+17$^{\circ}$4708  &sdF8    & ~9.47  &2 &1-6 &3-6 \\
P330E       &G2V     & 13.03  &3 &1  &1-5 \\
KF06T2      &K1.5III & 13.8   &3 &0  &3 \\
VB8         &M7V     & 16.70  &4 &3  &3 \\ 
2M0036+18   &L3.5    & 21.34  &1 &2  &2 \\
2M0559-14   &T4.5    &I=19.14 &2 &1  &1 \\
\enddata
\tablenotetext{(a)}{Simbad}
\tablenotetext{(b)}{GRW$+70^{\circ}$5824 is the STIS UV monitor standard, which is well
observed below 3062~\AA\ but is poor at longer wavelengths with only 1--2
observations.} \label{table:stars} \end{deluxetable}

Table~\ref{table:stars} lists the HST standard stars utilized for the flux
calibration; and the
three primary hot WD stars G191B2B, GD153, and GD71 define the ACS flux
calibration and its change over time. Multiple observations of these stars
have a sub-percent repeatability and provide a robust measure of the
cross-calibration between the Space Telescope Imaging Spectrograph
(STIS) and the ACS average response in each filter. For
the secondary standards, three separate ACS visits and three STIS measures of
their spectral energy distributions (SEDs) are required for each star to provide
minimal confirmation of repeatability and $<<$1\% statistical significance.
Results based on stars with fewer observations are less reliable. The measured
STIS SEDs are from CALSPEC \citep{bohlin14}, while ACS data obtained after 2011
July are not previously analyzed. The standards are all observed at the center
of the ACS WFC1-1K subarray in order to provide a standard reference point and
anchor for the flat field calibrations. 

Since the previous analysis by \citep{bohlin2012}, the USEAFTER dates for some
superbias reference files were adjusted to be earlier. CCD detectors are
operated with a positive electronic bias level, so that noise or drifts in the
analog circuitry before digitization does not drive the measured charge to
negative levels. The pattern in the bias frames taken with zero exposure time
changes slowly and must be monitored as a function of time.
A USEAFTER date is the time when a reference file first becomes relevant.
For example, the bias
observations that occurred 4 days AFTER the G191B2B observation on 2011Nov10 now
have a USEAFTER of 2011Oct27, so that the bias correction is closer in time than
the previously used reference file with a USEAFTER of 2011Aug 25. While the
change in the 1\arcsec\ photometry is negligible ($<$0.02\%), the bias is
important for the infinite (5.5\arcsec) aperture, which changes by up to 0.6\%.
The EE in the infinite aperture defines the diffuse source calibration, but the
point source flux calibration in the smaller apertures is unaffected, as long as
the same EE factor for conversion from the smaller standard radius to infinite
is used to convert back to the smaller radius. Perhaps, the absolute calibration
PHOTFLAM header keyword for infinite aperture should be augmented to include the
small aperture EE keyword values for convenience and to avoid erroroneous
usage of obsolete EE values.

\section{Data Processing}	

Standard practice for ACS pipeline data processing is to obtain and subtract a
CCD bias image, which consists of data obtained within the same CCD anneal cycle
(currently every four weeks). Data fetched from the Mikulski Archive for Space
Telescopes (MAST) shortly after a program execution should always be
re-extracted from the archive after the final bias and dark reference files are
delivered with their usual 1--2 month delay. Before the ACS electronics repair
in 2009 April on SM4 that included an Application-Specific Integrated Circuit
(ASIC) to read the CCD data, the subarray bias could be extracted from the full
frame biases; but different clocking speeds in the ASIC for subarrays now
mandates separate subarray bias observations. Full frame WFC bias reference
files have a regular two-week cadence; but the post-SM4 subarray bias
observations occur only as needed in association with the science observations.
Originally, a regular cadence of post-SM4 WFC1-1K subarray bias observations
executed through 2012 Oct, but calibration programs must now schedule any
required bias frames along with the external observation. There are no WFC1-1K
bias observations with USEAFTER dates between 2012 Oct 14 and 2014 Sep 3. Of the
cooler stars used in this work, the only observations that lie in this  WFC1-1K
bias desert are for BD+17$^{\circ}$4708 on 2013 Sep 4 and VB8 on 2013 Mar 11.

Before the proper associated bias data were reduced to reference files, some of
the WFC1-1K subarray observations were processed with a bias reference file that
was obtained up to 666 days earlier. Aperture photometry of 1\arcsec\
and smaller is the same to 0.1\% for current and out-of-date bias reference
files. Even though individual measures with the 5.5\arcsec\ radius have
errors of up to 1\%, the effect on the average EE tabulated in the Apendix is
less than 0.2\% for any filter.

For the flat field mapping program 13602 \citep{bohgrog}, the 400x400 subarrays
have associated bias observations without overscan columns. Unfortunately, the
pipeline data processing requires bias observations with overscan, and the
archival 13602 data are processed with the bias data extracted from a full frame
bias image. The errors in the 20 pixel radius photometry for the archival MAST
versions range up to 1.7\% in comparison to the custom reduction of
\citet{bohgrog}.

In order to verify consistency with the various community photometry packages,
Table~\ref{table:data} includes the extracted $electrons~s^{-1}$ used in the
current flux calibration for aperture radii of 3, 5, 10, 20, and 110 pixels for
the WFC1-1K subarray observations of GD71 obtained at 2003.84 on 2003 Nov 3. The
images are \textit{j8v602*\_crj.fits} files, as processed by the CALACS pipeline on
2015 Nov 17 and then corrected by the pixel area map (PAM). Changes from the
same 20 pixel 1\arcsec\ photometry shown in table 3 of \citet{bohlin2012} are
$<0.06\%$. Changes in the sky backgound are similarly neglibible but do become
more consistent with a range of -0.024 to 0.094 vs. the previous -0.011 to 0.221
$electrons~s^{-1}$.

\begin{deluxetable}{cccrrrrrrr}     
\tablewidth{0pt}
\tablecolumns{10}
\tablecaption{Sample WFC photometry in $electrons~s^{-1}$ for GD71 from 
j8v602*crj.fits files, as corrected by the PAM reference file.}
\tablehead{
\colhead{Filter} &\colhead{x (px)} &\colhead{y (px)} 
&\colhead{exp (s)}&\colhead{3 px} &\colhead{5 px}
&\colhead{10 px} &\colhead{20 px} 
&\colhead{110 px} &\colhead{sky} }
\startdata
F435W&  497.8&  514.5&   4.0& 127926& 140928& 147862&  153310& 163968& -0.007 \\
F475W&  497.7&  514.9&   3.0& 162872& 180284& 189444&  196187& 210275& -0.024 \\
F502N&  497.3&  514.9&  90.0&	4660&	5164&	5433&	 5622&   5987&  0.000 \\
F555W&  497.6&  514.6&   4.6&  97517& 108059& 114029&  117966& 125763&  0.013 \\
F550M&  497.6&  515.2&  10.0&  41260&  45694&  48391&	50188&  53620&  0.001 \\
F606W&  497.4&  514.6&   2.6& 171574& 190340& 201716&  208721& 220635&  0.094 \\
F625W&  497.4&  514.9&   5.0&  86432&  95947& 102064&  105772& 112404&  0.017 \\
F658N&  497.9&  515.2& 120.0&	3832&	4263&	4555&	 4737&   5013&  0.001 \\
F660N&  498.0&  514.5& 300.0&	1510&	1666&	1780&	 1877&   1984&  0.000 \\
F775W&  497.4&  514.7&   9.0&  47405&  53338&  57179&	59280&  63064&  0.011 \\
F814W&  497.9&  514.5&   8.0&  56248&  64280&  69303&	71881&  75904&  0.004 \\
F892N& 1009.2& 1025.0& 220.0&	1638&	1955&	2139&	 2221&   2383&  0.001 \\
F850LP&  497.6&  515.1&  24.0&  14872&  18050&  20056&  21079&  22641&  0.001 \\
\enddata
\label{table:data}
\end{deluxetable}

\section{Encircled-Energy Aperture-Correction Update}	

\subsection{Equations}

Following the methodology of \citet{bohlin2012} but with the improved notation
of \citet{bohlinetal14}, the photon weighted mean flux over the bandpass in
wavelength ($\lambda$) units is  \begin{equation}\langle F\rangle={\int
F_{\lambda}~\lambda~R~d\lambda \over \int
\lambda~R~d\lambda}\label{fav}={S~N_e}~,\end{equation}  where our flux $F$ is
$erg~cm^{-2}~s^{-1}~\AA^{-1}$, $R$ is the system  fractional throughput, i.e.,
the total system quantum efficiency (QE) as a function of wavelength, S is the
instrumental calibration constant \textit{photflam} that appears in the ACS data
headers, $N_e$ is the measured instrumental count rate in photoelectrons
$s^{-1}$ in the ACS  'infinite' 5.5\arcsec\ radius aperture, and the integral is
over the bandpass range of the filter. Equation 1 includes all the photons from
starlight incident on the primary, regardless of where the light hits the
detector, so that $N_e$ implicitly includes all the detected photons in an
infinite-radius aperture. Of course, an actual infinite radius is not feasible
and 5.5\arcsec\ is about as large an aperture that is practical. In principle,
models of the instrumental point spread function  (PSF) could help define the
encircled energy for an infinite aperture. In the case of HST, considerable
effort has been expended on the Tiny Tim PSF modeling software. However, the
Tiny Tim user manual states: ``Generally, the models are not very good past a
radius of $\sim$2\arcsec, due to the effects of scatter and high-frequency
aberrations'' \citep{krist04}. Because of the small gains in enclosed signal as
the aperture size increases toward 5.5\arcsec, the error in the infinite
aperture estimate should be $<$1\%; and this uncertainty applies only to the
diffuse source calibration. The precision of the point source flux calibration
is not affected by errors in the adopted infinite aperture EE estimate. The sky
background annuli are 6-8\arcsec\ for  WFC and 5.6-6.5\arcsec\ for HRC. The
count rate $N_e$ can also be predicted from the known instrumental throughput
parameters and the known CALSPEC flux  $F_\lambda$ for each standard star 
\begin{equation}N_e={A \over hc}{\int F_{\lambda}~\lambda~R~d\lambda}~,
\end{equation} so that  \begin{equation}S={hc \over A\int{\lambda~R~d\lambda}}~,
\label{pl}\end{equation} where h is Planck's constant, c is the speed of light,
and A is the collecting area of the primary mirror, taken to be 45,239 $cm^2$.
The traditional method of flux calibration for HST that is followed in this
paper involves adjusting $R$ in Equation 2, so that the computed $N_e$ equals
the measured count rate. Systematic deviations between the computed and measured
$N_e$ for all the filters are accounted by smooth adjustments of the detector QE
vs. wavelength, while residuals for each filter are corrected by adjusting the
filter throughput. The point source calibration $S$ for the infinite aperture is
required for computing the diffuse source surface brightness calibration
\citep{bohlinetal14}.

For aperture photometry with a radius of $i$, the EE fractions $E_i$ are the
average of the measured $E_i=N_i/N_e$ for each filter, i.e. the sum of the
counts $s^{-1}$ $N_i$ within a radius $i$ divided by $N_e$; and once the average
$E_i$ are established, the infinite aperture $N_e$ can be computed from any
small aperture photometry by dividing by the encircled energy fractions
$E_i$ that are the aperture corrections used to correct small radius aperture
photometry to a larger reference aperture, i.e. the infinite aperture for ACS.

For example, the commonly used 3, 5, 10, and 20 pixel radii for WFC (double 
for HRC) are approximately 0.15, 0.25, 0.5, and 1 arcsec, where
\begin{equation}N_e
= N_3/E_3 = N_5/E_5 = N_{10}/E_{10} =  N_{20}/E_{20}~.\end{equation}
Thus, the mean flux $\langle F\rangle$ is S~N$_i/E_i$, where the most 
appropriate radius subscript can be chosen. 

For many science images, the use of the proper sky annulus at 6--8\arcsec\ is
not appropriate and a correction is required to the photometric calibrations
presented here that are all based on that fixed sky annulus. Ignoring these
corrections for sky annulii of only a few pixels causes photometric errors of a
few percent. Because of gradients in the sky background in many science images
from the star to the WFC standard sky annulus at 6--8\arcsec, smaller sky annuli
are often required. However, the stellar signal from the PSF wings that are
included in the sky measures must be added to the measured count rate $N_i$. The
fraction $f$ of the signal in the sky annulus depends on the photometry radius
$r_i$ and the inner $r_a$ and outer $r_b$ sky radii. The measured count rate
$N_i$ must be divided by $(1-f)$ to get the true photometry $N_i)$, where
\begin{equation}f={{E_b/E_i - E_a/E_i} \over {{r_b}^2/{r_i}^2 -
{r_a}^2/{r_i}^2}}~.\end{equation} 
For example, three pixel WFC photometry  $N_3$
with $r_i=3$, background inner annulus $r_a=4$, and outer $r_b=6$  has a
correction $(1-f)=0.963$ for F850LP according to the EE tables in the Appendix.

\subsection{EE Changes over Time}   

\begin{figure}	
\centering 
\includegraphics[height=8in]{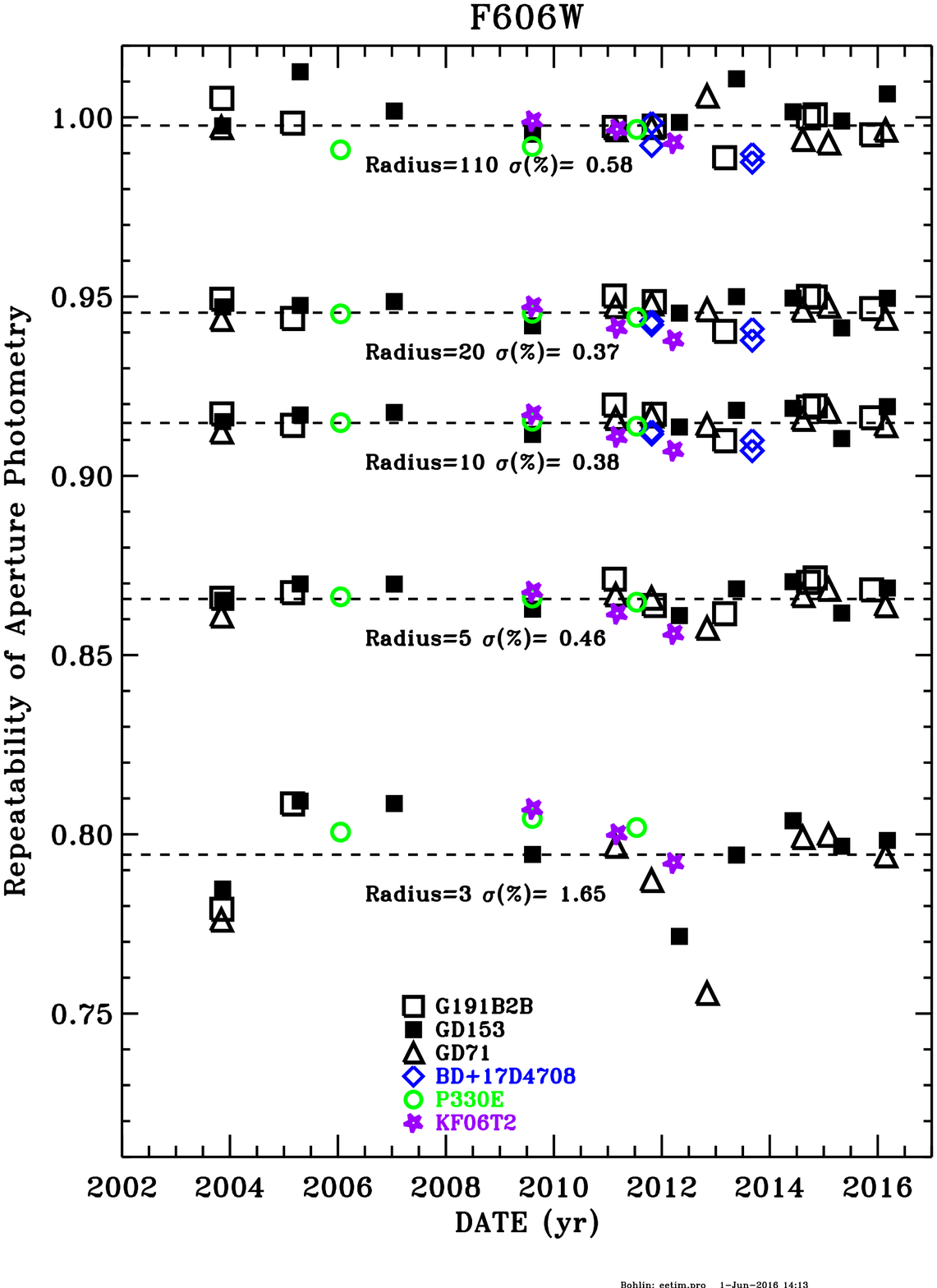}
\caption{\baselineskip=14pt
Ratio of the measured WFC count rates Ni to the predicted count rate in an
infinite aperture for G191B2B, GD153, GD71, BD+17$^{\circ}$4708, P330E, and
KF06T2. The gap in time from 2007-2009 is when ACS/WFC was not operational. The
low point for the three pixel radius GD71 at 2012.8 is not caused by any
saturated pixel falling outside the three pixel radius, which suggests that the
occasional image has a 3$\sigma$ error (5\% here) for three pixel
photometry.   \label{ee606}} \end{figure}

Plots of the aperture photometry vs. time illustrate the reproducability and
determine the best aperture for the absolute flux calibration. For example,
Figure~\ref{ee606} compares the measured WFC F606W count rate $N_i$ to the
predicted $N_e$ for an infinite aperture for five of the most relevant
photometry apertures from three to 110 pixels. The measured photometry for each
star is divided by the predicted value in order to preserve the intrinsic
rms scatter of the data and avoid large fluctuations that would be caused by
dividing by the measured infinite aperture values. The rms scatter is tabulated
for all eight of the WFC broadband filters in Table~\ref{table:sig}. There are
no obvious trends with time or with stellar color in Figure~\ref{ee606}.

\begin{figure}	
\centering 
\includegraphics[height=8in]{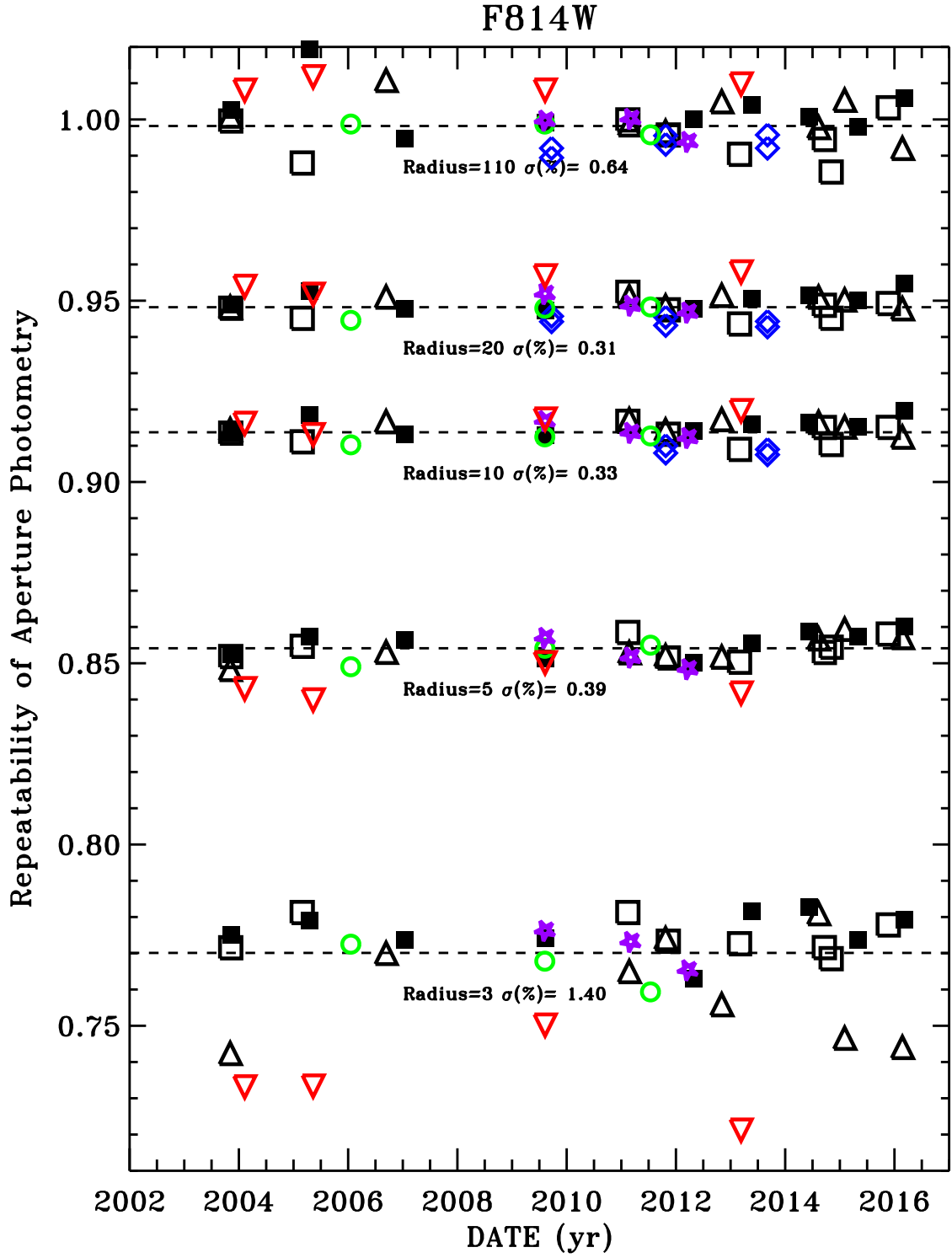}
\caption{\baselineskip=14pt
As in Figure~\ref{ee606} for F814W and with the addition of VB8 (red inverted
triangles).
\label{ee814}} \end{figure}

\begin{figure} 
\centering  
\includegraphics*[height=7.0in]{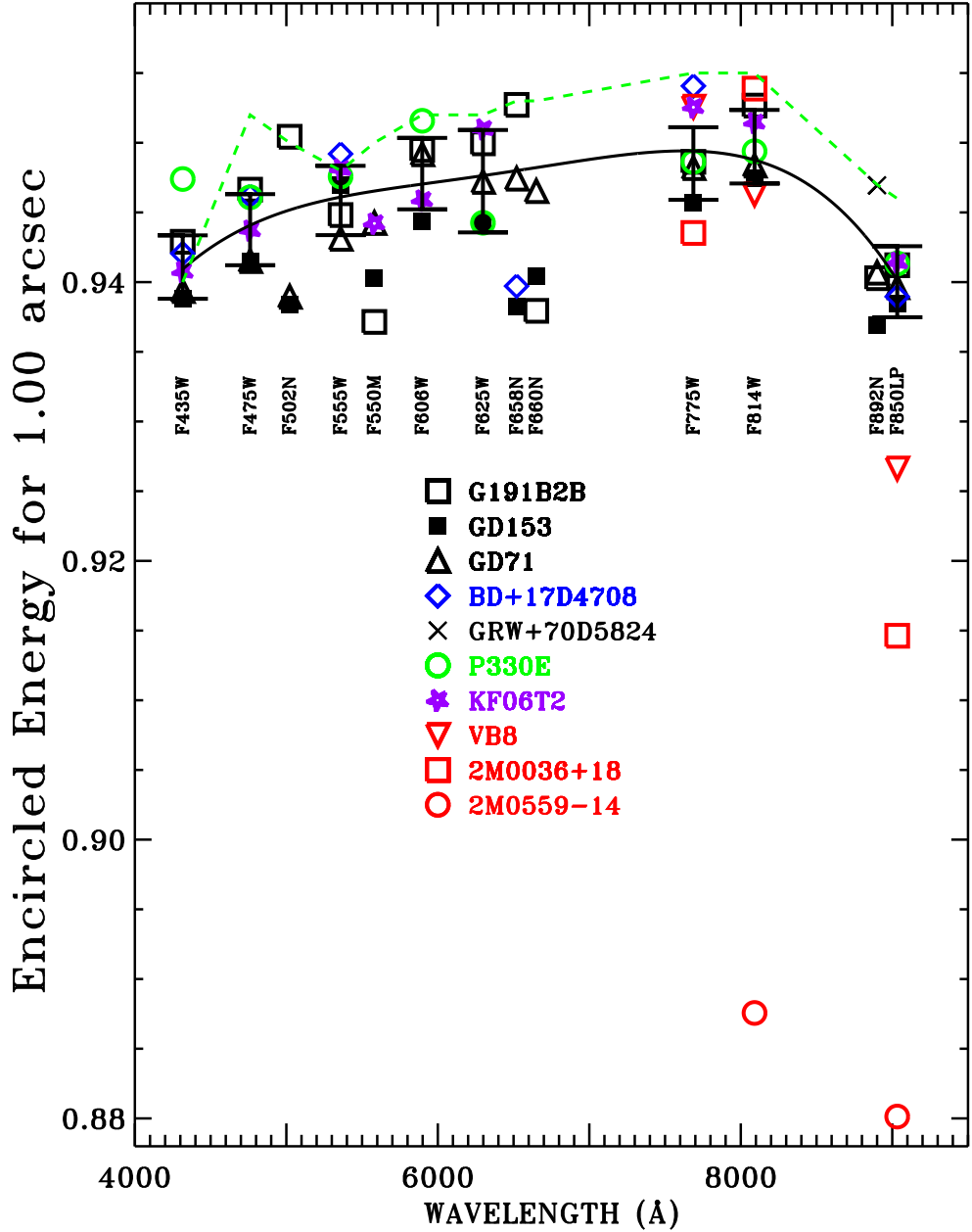} 
\caption{\baselineskip=12pt
Encircled energy $E_{20}$ for WFC, i.e. fraction of signal in a 1\arcsec\ aperture
relative to an ``infinite'' aperture of 5.5 arcsec radius. The solid black line
is a quartic fit to only the broadband filter EE averages for the WDs, the F star
(BD+17$^{\circ}$4708), the G star (P330E), and the K star (KF06T2). Error bars
for the eight filters used for the polynomial fit
are $\pm3\sigma$ errors-in-the-mean of these averages, but the average values 
are not shown to avoid crowding. The green dashed lines are the 
corresponding results of \citet{sirianni05}. 
\label{eewfc}} \end{figure}

\begin{deluxetable}{cccccc}     
\tablewidth{0pt}
\tablecolumns{6}
\tablecaption{Photometric rms Scatter for Various Radii at the WFC1-1K Reference Point}
\tablehead{
\colhead{Filter}
&\colhead{$\sigma_{3}$(\%)}
&\colhead{$\sigma_{5}$(\%)} &\colhead{$\sigma_{10}$(\%)}
&\colhead{$\sigma_{20}$(\%)} &\colhead{$\sigma_{110}$(\%)}
}
\startdata
F435W   &1.00 &0.43 &0.37 &0.35	 &1.08 \\
F475W   &1.76 &0.39 &0.31 &0.30 &0.51 \\
F555W   &1.26 &0.41 &0.41 &0.41 &0.69 \\
F606W   &1.65 &0.46 &0.38 &0.37 &0.58 \\
F625W   &2.00 &0.43 &0.30 &0.30 &0.77 \\
F775W   &2.23 &0.60 &0.50 &0.51 &0.76 \\
F814W   &1.40 &0.39 &0.33 &0.31 &0.64 \\
F850LP  &1.94 &0.87 &0.58 &0.44 &0.81 \\
\enddata 
\label{table:sig}
\end{deluxetable}
For the bright, isolated standard stars, the 20 pixel ($\sim$1\arcsec) aperture
generally has the best repeatability at 0.30--0.51\% for $\sigma_{20}$ and, therefore,
defines the best photometry for each observation, although there is little loss
in precision with an aperture as small as five pixels, where
0.39$<\sigma_5<$0.87. While the infinite 5.5\arcsec\ radius aperture contains
all the signal by definition, there is more scatter because of the excess sky
noise in such a huge aperture. The three pixel radius aperture has up to seven
times the uncertainty of a 1\arcsec\ aperture; and if possible, the use of $E_3$
to calibrate science data should be avoided because of exact centering
uncertainties and focus variations that could be systematic over an entire
image.

Figure~\ref{ee814} is similar to Figure~\ref{ee606}, except for the addition of
the problematic M star VB8. The infinite (110 pixel) radius values for all four
VB8 observations are high by 0.5--1\% with respect to the hotter star average.
This offset could be caused by a low value of the predicted $N_e$ from the STIS
SED. In the region of the F814W sensitivity, the VB8 flux is on the short
wavelength, rapidily declining side of the SED, where stellar variability is
common. In addition, a large contribution to the VB8 predicted count rate is
from 9000--9600~\AA\, where STIS has a low sensitivity and worse rms
repeatability. In Figure~\ref{ee814}, the 20 pixel radius VB8 points are also
high but not as high as the 110 pixel values, which accounts for the slightly
low F814W, VB8 average for the WFC in Figure~\ref{eewfc}. At the three pixel
radius in Figure~\ref{ee814}, VB8 is significantly low, indicating that there
must be some extra loss into the red halo of Section 4.4 for such a small
aperture.

\subsection{Hot, Early Type Stars}		

\begin{figure} 
\centering  
\includegraphics*[height=7.0in]{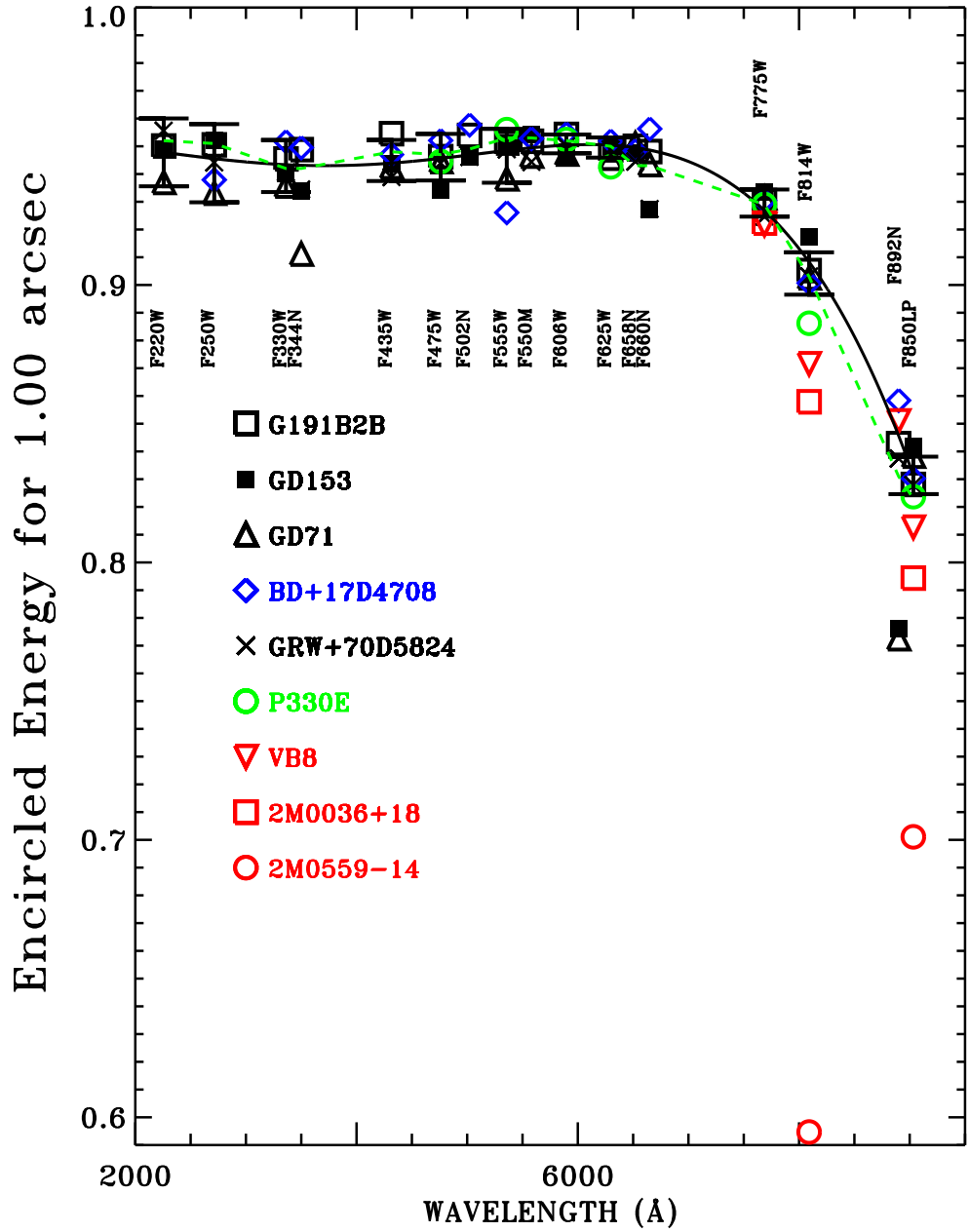}
\caption{\baselineskip=12pt
As in Figure~\ref{eewfc} for an HRC 1\arcsec\ aperture, except that there  are
no data for the K star (KF06T2) and the quartic fit is to the 11 broadband
filters that are shown as error bars. \label{eehrc}} \end{figure}

EE values are the measured $N_i$ divided by the noisy measured $N_{110}$ values;
and fitting techniques are required to lower the uncertainties. While the large
110 pixel infinite aperture should recapture all the trapped and re-emitted
charge, the smaller aperture data require a CTE correction. The pipeline
processing now supplies the \citet{Anderson2010} CTE corrected full frame
4096x4096 pixel images; but CTE corrected subarrays are not provided, so that
the WFC1-1K subarrays used here must be corrected for CTE losses in the
post-pipeline processing by the method of \citet{bohlinjay2011}. A fit as a
function of wavelength for the broad filters of Table~\ref{table:sig} reduces
the uncertainty, while at the same time providing fitted values for the
problematic narrower filters \citep{bohlin2012}. Figures~\ref{eewfc} and
\ref{eehrc} are examples of polynomial fits to the average EE as a function of
the pivot wavelength for the 1\arcsec\ aperture from the hot stars to the
K1.5III KF06T2. The plotted symbols are the averages for each star; but the fit
is to the global averages of all of the individual observations for each
broadband filter. The final results of similar fitting for 12 aperture sizes
appear in Table~\ref{table:eewfc} and Table~\ref{table:eehrc} in the Appendix,
where typical uncertainties in the EE fits for WFC and the 0.15, 0.25, 0.5, and
1 \arcsec\ radii are 0.4, 0.2, 0.4, and 0.1\%, respectively. Correspondingly for
HRC, the respective uncertainties are  0.4, 0.3, 0.4, and 0.3\%. The
\textit{\_flt.fits} files must be used for the sometimes saturated WFC data for
the bright  BD+17$^{\circ}$4708, because saturated pixels cannot be combined
with the cosmic ray rejection algorithm used to produce the \textit{\_crj.fits}
files. These final EE values are fits to the measured ratios of the small to
infinite aperture count rates, where a few of the later single exposure
\textit{\_flt.fits} BD+17$^{\circ}$4708 observations in the big 5.5\arcsec\
aperture deviate from the mean by $>3\sigma$ and are omitted. The ACS CCDs are
linear through saturation, as long as  all of the saturated signal is included
in the aperture \citep{gillil2004}, which means that some of the smaller
aperture measures of BD+17$^{\circ}$4708 must also be omitted.  The maximum
deviation from \citet{sirianni05} for the broadband WFC filters is for the three
pixel F435W EE value of 0.792, whereas \citet{sirianni05} has 0.775.

\subsection{Cooler, Late Type Stars}	

For stars redder than KF06T2 (K1.5III), the flux calibration does not achieve
the goal of 1\% accuracy because of the uncertain EE from the
sparsely observed and possibly variable M, L, and T stars. Red stars of spectral
type M and later have a ``red halo'' and show significantly lower EE, because
long wavelength photons scatter more in the CCD substrates, especially for HRC,
which lacks the special anti-scattering layer incorporated into the WFC CCDs.
Figures~\ref{eewfc} and \ref{eehrc} show these red star EE values as red data
points. \citet{sirianni05} provide their tables 6-7, which define the EE
vs.~effective wavelength. These effective wavelengths are not constant for a
filter but increase as the peak of the stellar flux distributions move to longer
wavelengths. There is little new data for the three cool stars, so the
discussion of uncertainties in \citet{bohlin2011,bohlin2012} still stands with
error bars of $\sim$4\% for the \citet{sirianni05} EE results for stars of type
M and later that are recommended for the WFC filters F814W, F850LP, and
F892N and for the HRC filters F775W, F814W, F850LP, and F892N. The only
exceptions are for WFC F814W and F892N for the reddest star (T6.5) 2M0559-14,
where 0.89 instead of the \citet{sirianni05} EE values of 0.94 is recommended
for the 20 pixel aperture and 0.84 instead of 0.90 for the 10 pixel aperture.
For completeness, some of the most commonly used and recommended
\citet{sirianni05} EE values for cool stars are in Table~\ref{table:eered} for
EE values that differ from those in the Appendix for stars of K type and hotter.
There are EE values for more radii in \citet{sirianni05}.

\begin{table}[!h]	
\begin{center} 
\begin{tabular}{|c|c|c|c|c|c|c|}
\hline

\multicolumn{1}{|c}{Filter} & 
\multicolumn{2}{|c}{VB8 (M7)} & \multicolumn{2}{|c}{2M0036(L3.5)} &
\multicolumn{2}{|c|}{2M0559(T6.5)} \\

& \multicolumn{2}{|c}{10px~~~~20px}  &
\multicolumn{2}{|c}{10px~~~~20px} &
\multicolumn{2}{|c|}{10px~~~~20px} \\

\hline \hline

\bf WFC     &       &         &	 & 	    &        & 	\\
F814W  &  ...   &  ...  &  ...   &  ... & 0.84* & 0.89*  \\
F892N  &  ...   &  ...  &  ...   &  ... & 0.84* & 0.89*  \\
F850LP & 0.87   & 0.93  &  0.85  & 0.92 & 0.78  & 0.88   \\
\bf HRC    &       &         &	 & 	     &        &  \\
F775W  & 0.87   & 0.91 &   0.86   & 0.91  & 0.85  & 0.90  \\
F814W  & 0.79   & 0.86 &   0.78   & 0.85  & 0.75  & 0.82  \\
F892N  & 0.77   & 0.84 &   0.77   & 0.84  & 0.77  & 0.84  \\
F850LP & 0.69   & 0.78 &   0.67   & 0.76  & 0.55  & 0.66  \\
\hline
\end{tabular}
\end{center}

\caption{\textsl{Encircled energy aperture corrections for very cool stars
	from \citet{sirianni05}, except those marked with an asterisk, which are
	from this work. Use the Appendix EE values for cases not specified here.}} 
\label{table:eered}
\end{table}

\subsection{EE Variations around the Field of View}	

\citet{bohgrog} investigated the flat field variation for F435W and F814W with
two stars at positions spaced around the WFC field-of-view (FOV); and those same
data reveal any variation in EE as a function of field position. While the flat
fields plus PAM files correct the stellar signal strength to a uniform value
around the  field in the \textit{\_crj.fits}, there is no correction of the EE
for the pixel size or focus variations. Thus, the EE fraction may vary,
expecially for small aperture radii. With respect to the Table~\ref{table:eewfc}
EE at the WFC1-1K reference point on CCD chip-1, the difference in the EE is
measured at the field positions of \citet{bohgrog} for the 3, 5, 10, and 20
pixel apertures. The expected 1$\sigma$ differences are from
Table~\ref{table:sig}. For example for the average of the two stars, the
uncertainty in $E_{20}$, i.e. the 1\arcsec\ 20 pixel radius divided by the
infinite 110 pixel photometry is $\sigma=\sqrt{(0.35^2+1.08^2)/2}=0.8$\% for
F435W. The only 3$\sigma$ deviations are +2.0--3\% larger EE for four of the
F814W smaller pixel radii near the (3839,1791) location in the 4096x4096 FOV,
eg. 0.873 $\pm0.005$ instead of the 0.853 $\pm0.002$ EE in
Table~\ref{table:eewfc} for a five pixel radius. However, there are several
+2$\sigma$ deviations near the same location in the upper right hand corner of
chip-2 for both F435W and F814W. Thus, systematic differences in EE around the
field are possible with a range of $\pm$1\% for 20 pixels to $\pm$2\% for five
pixel radius and $\pm$3\% for three pixel radius.

In summary, the best chance of achieving the 1\% photometry goal is to place the
target star at the well characterized WFC1-1K reference point, where several
measures of several stars reduce the uncertainties in the average EE to
0.1--0.3\%.

\section{\bf Changes in Sensitivity with Time} 

In order to specify a flux calibration for the ACS CCD imaging data, the
photometry must be corrected for gradual losses in the instrumental throughput 
since the installation of ACS into HST on 2002 March 7 (2002.16). The side-1
electronics failure occurred on 2006 Jan 27 \citep{bohlinetal11}; and the
activation of the  side-2 electronics with the reduction of the set-point
temperature from -77C to -81C for the WFC CCD on 2006 July 6 (2006.50)
introduced a discontinuity in sensitivity \citep{mack07}. In addition, the
lowering of the set-point to -81C caused minor flat field changes of up to
0.6\%, which are automatically included in the STScI ACS pipeline data products 
\citep{gilliland06}. A second discontinuity in sensitivity is due to the CCD
Electronics Box Replacement (CEB-R) with its ASIC sidecar circuit board during
the HST Servicing Mission SM4 in 2009 May at 2009.4 \citep{grogin10}.

This work updates the results of \citet{bohlinetal11}, where more details about
the analysis are provided. The reference positions for the flux calibration are
at the center of the HRC CCD and at the center of the standard WFC1-1K subarray
at pixel (3583,1535) on chip-1; and the change in sensitivity is determined by
the photometry with the best repeatability, i.e. the 1\arcsec\ $N_{20}$
photometry of the three HST primary WD standard stars G191B2B, GD153, and GD71.
The sub-percent CTE corrections of \citet{bohlinjay2011} are applied for WFC and
should be negligible for the early HRC data, which is heavily exposed. The three
stars are on a common scale defined by their observed infinite-aperture count
rate C=$N_{20}/E_{20}$ divided by the predicted total count rate $P=N_e$ from
the synthetic photometry of Equation 2. The P values depend only on the stellar
flux and the system throughput for the filter and are corrected only for the 
discontinuity due to the change in the WFC temperature from -77C to -81C. The
stellar SEDs for these three fundamental HST standards are the measured STIS
fluxes on the HST scale \citep{bohlin14} from the HST CALSPEC
database\footnote{http://www.stsci.edu/hst/observatory/crds/calspec.html/}. 

\begin{figure} 
\centering  
\includegraphics*[height=8.0in]{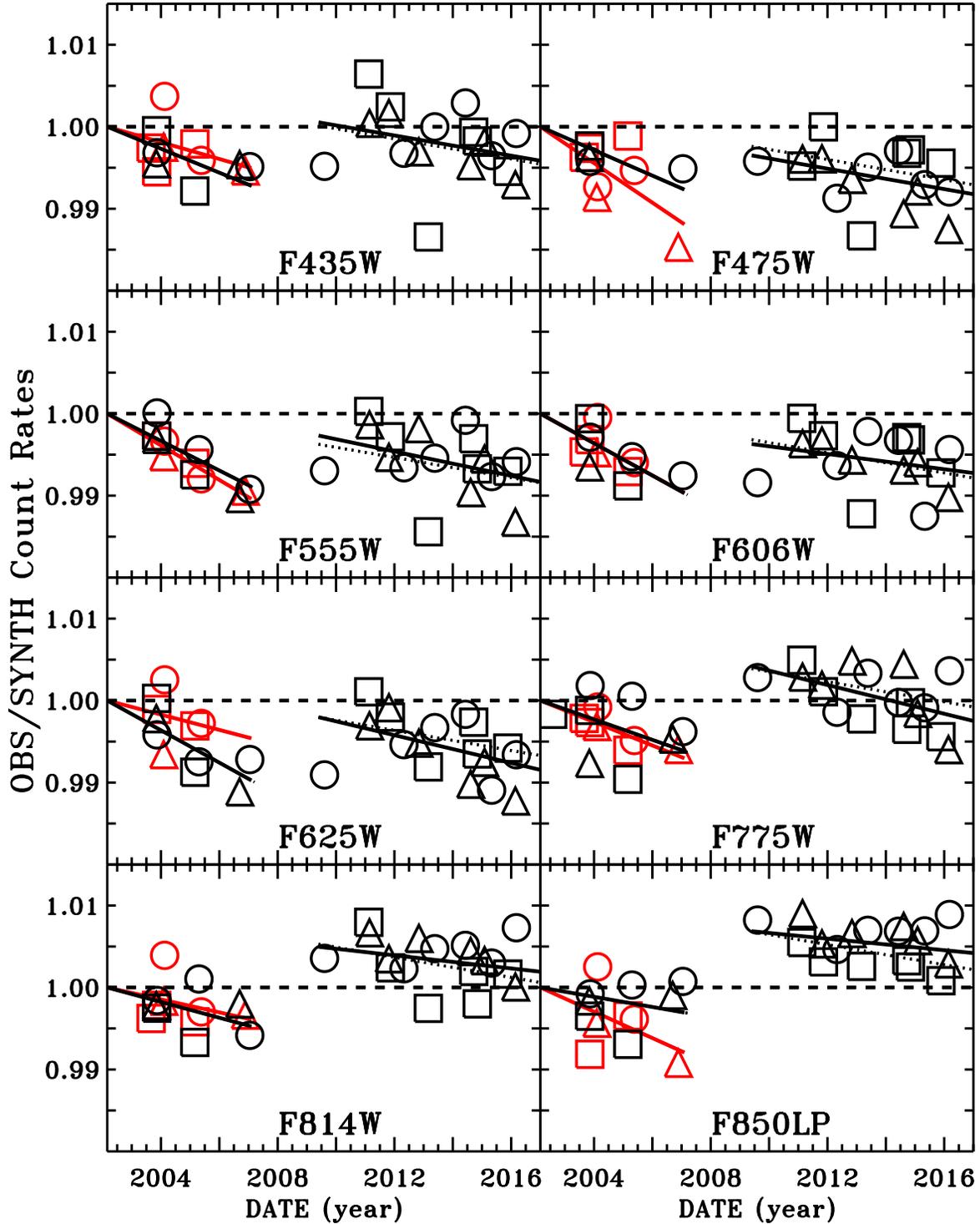}
\caption{\baselineskip=12pt
Ratio of observed one arcsec radius photometry to predicted count rates for 
eight broadband filters as a function of time. The observed count rate is
corrected for EE and CTE. Symbols are: square-G191B2B, circle-GD153, and
triangle-GD71. Black symbols are for WFC, and red are HRC. The solid line fits
and all of the data for  each camera are normalized to the value which makes
each fit equal to unity at 2002.16. The dotted lines are the parameterized fits
used by the routine $acs\_timecorr.pro$, where after  2009.4, the slope is set
to the average of -0.061\%/yr; and the value at 2009.4 is from
Figure~\ref{tchpost}. For the pre-SM4 WFC, the
dotted and solid lines coincide. \label{tchfilt}} \end{figure}

\begin{figure} 
\centering  
\includegraphics*[height=5.in]{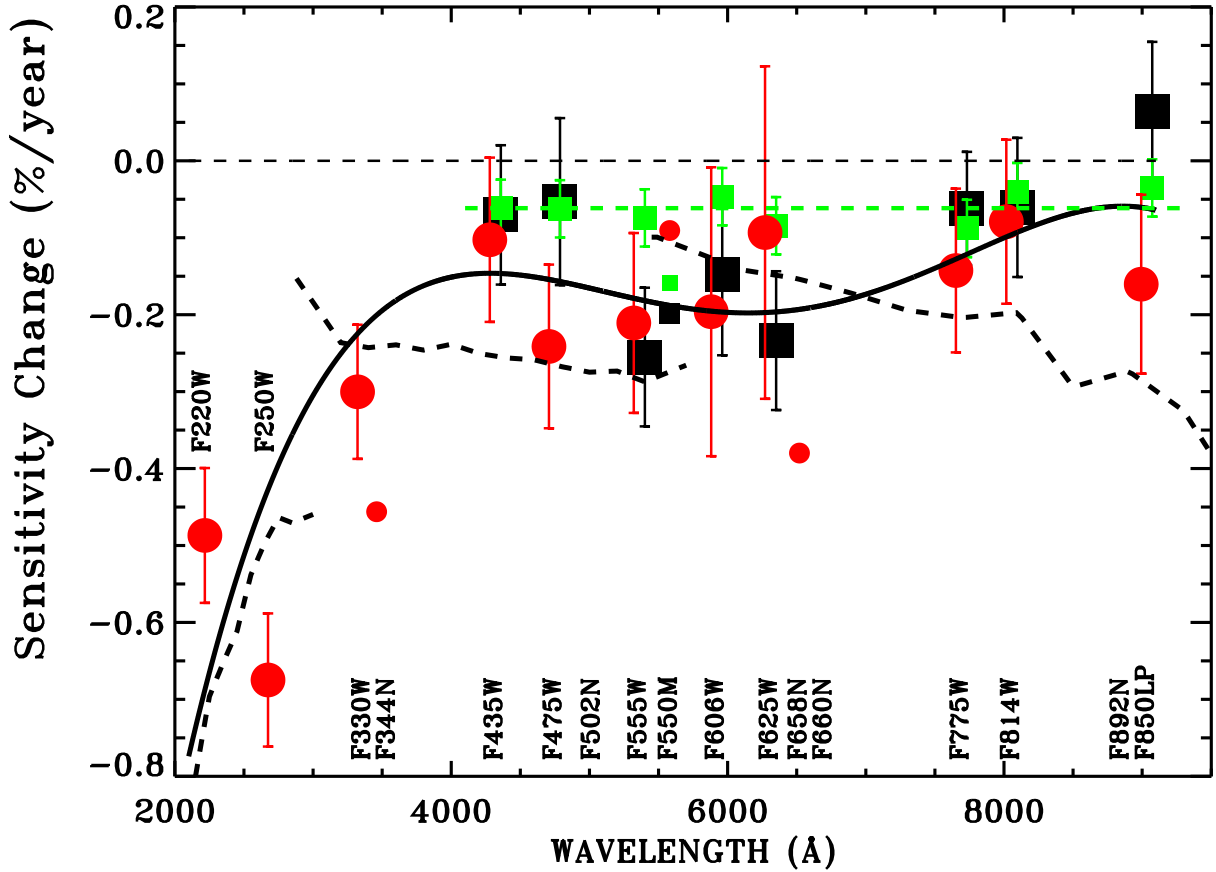}
\caption{\baselineskip=12pt
Percent loss of sensitivity ($100{\Delta}S$) per year for all filters, as in
Figure~\ref{tchfilt}. Black is for the pre-SM4 WFC, red represents
the pre-SM4 HRC, and green is the post-SM4 WFC. Large symbols with error bars
are broad bandpass, while small symbols are for the narrow and medium  band
filters. A fourth order fit (solid black line) to the large pre-SM4 symbols is
shown as a function of the filter pivot wavelength, while the green dashed line
is the average post-SM4 loss rate. Error bars are the 1$\sigma$ formal
uncertainty in the slope of the linear fits. All but three of the error bars
cross the fits. The black dashed lines are the average sensitivity losses over 
the 2002-2007 time frame for
the three CCD grating modes of STIS \citep{stys2004} and are similar to the ACS
loss rates. Because an increase in sensitivity seems non-physical, the pre-SM4
fit for F850LP is forced to the more robust post-SM4 average. \label{tchpre}}
\end{figure}

For the first years of ACS operations, the change in response that is measured
by C/P for each filter is fit with a straight line; and then to reduce
measurement noise and enforce the prior of the expected smooth change with
wavelength, these line slopes are fit with a polynomial as a function of the
filter pivot wavelengths as shown in Figure~\ref{tchfilt} for the broadband
filters available to both WFC and HRC. The HRC and WFC photometry are in red and
black, respectively, along with their separate linear fits. The P values for WFC
are corrected for the first sensitivity discontinuity for the switch of the
set-point temperature from -77C to \mbox{-81C} at 2006.5 \citep{mack07} but are
NOT adjusted for the second discontinuity at 2009.4 for the SM4 repair. While
the HRC red solid lines are the actual linear least-square fits to the data, the
WFC pre-SM4 fits are for the constant term only and adopt the final fitted
slopes from Figure~\ref{tchpre}, in order to get the best estimate of the
post-SM4 normalization level. The WFC fits generally appear to agree with the
data within the rms noise. However for F850LP, the positive slope of 0.06\%/yr
from Figure~\ref{tchpre} and Table~\ref{table:losses} is clearly a better fit to
the minimal WFC data.

For the pre-SM4 ACS epoch, the data are rather sparse, because the policy of the
once per year monitoring of the sensitivity changes at the WFC1-1K reference
point with the three primary standard stars had not yet been established. The
filter F814W is typical of the better data sets with seven WFC data points that
agree well with the seven HRC observations, as shown in Figure~\ref{tchfilt}.
F475W is a worst case, where there are only four WFC data points; and the main
constraint on the rate of sensitivity loss is the lone GD153 point at 2007.0
from program 11054. The planned observations of G191B2B in that cycle 15 were
not executed because of the ACS failure at 2007.1. Given the scatter in the
data, the WFC loss rate is uncertain but would coincide with the
slope from the fitted results, if the lone 2007.0 point that has all the
leverage on the fit were $\sim$1$\sigma$ lower. Furthermore, the lone HRC point at
2006.9 for GD71 is probably a low statistical outlier. Statistically, the best
result is from accepting both the WFC and HRC measures and adopting the fitted
mean for F475W. 

\citet{bohlinetal11} present a figure 1 for F555W that is similar to F555W in
Figure~\ref{tchfilt},
but where the results for GD153 are systematically low by a fraction of a
percent. In particular, the GD153 observation at 2009.6 was below the fit by
0.7\%, while the revised results for F555W have this same data point with the
typical scatter of only 0.4\% low. The primary reason for this improvement among
all the analysis updates listed in the Introduction is that the new CALSPEC
fluxes for GD153 in the 5500\AA\ region are $\sim$0.4\% lower with respect to
the other two primary standards, which makes the denominator smaller and raises
that point to only $\sim$1$\sigma$ below the fit. Now, the F555W scatter is
typical of the other filters, and the \citet{bohlinetal11} anomaly is remedied.

For the later years of ACS after SM4 at 2009.4, the eight WFC broadband filters
are well monitored with $\sim$18 visits; and all show a consistent sensitivity
loss with a statistical significance ranging from 0.9 to 2.4$\sigma$. This
average post-SM4 loss rate is 0.00061 per year, i.e. 0.061\%/yr with an
error-in-the-mean of 0.007\%/yr, which differs at most by $<1\sigma$ from the
measured rate of 0.035 $\pm0.037$\%/yr for F850LP. The average loss rate of
0.061\%/yr is adopted for all WFC data after 2009.4. The general trends of
reduced loss rates with time and increased loss rates toward UV wavelengths are
the expected behavior for a gradually slowing outgassing of hydrocarbons and
their polymnerization on the optical surfaces. The observed trends in STIS
follow this model.

Following \citet{bohlinetal11}, Figure~\ref{tchpre} shows the slope of the rates
of sensitivity loss (${\Delta}S$) for each filter and a fourth-order polynomial
fit to the pre-SM4 data for both cameras as a function of the pivot wavelength
($\lambda$). 
\begin{equation}{\Delta}S=a+b\lambda+c\lambda^2+d\lambda^3+e\lambda^4
\end{equation} The medium and narrow band data are shown but are omitted in the
fitting procedure because of the minimal ACS observation sets and also because
of the finite STIS resolution that affects the predicted count rate P. The
F850LP WFC and HRC 1$\sigma$ error bars nearly touch, and the other seven
broadband pairs of 1$\sigma$ error bars all overlap. There are no systematic
differences between the two cameras; and thus, the HRC and WFC slopes are fit
together. Among the 16 data points for the eight pairs of WFC and HRC
broadbands, only the WFC F850LP lies more than 1$\sigma$ from the fourth-order
fit; and the 1$\sigma$ uncertainty of 0.08\%/year in the fit encompasses most of
the pre-SM4 measurements. The coefficients of the fourth order polynomial,
pre-SM4 fit in Figure~\ref{tchpre} are \mbox{a=-0.044561}, b=3.0376e-05, c=-7.7328e-09,
d=8.3760e-13, e=-3.2569e-17, while the loss rates evaluated at the average pivot
wavelengths appear in Table~\ref{table:losses} along with the measured values.
The fitted loss rates also appear in mag/year for comparison with the pre-SM4 47
Tuc results of \citep{ubeda13} for the WFC. The two entirely different methods
agree with a maximum difference in slope of 0.017--0.0018 mag/year for F475W,
F658n, and F660N, where \citep{ubeda13} have their largest uncertainties of
0.0015--0.018 mag/year. The main difference in the two methods is that
\citet{ubeda13} did not enforce the prior of a smooth variation of the loss
rates vs. wavelength. The STIS loss rates appear as dashed lines in
Figure~\ref{tchpre} and suggest that the ACS sensitivity losses should also be a
smooth function of wavelength. The overall level and not the slopes of the STIS
loci should be compared to ACS, because the STIS slope is affected by the
grating blaze shifts due to sublimation of the epoxy that bonds the replica
gratings to the subtrate.

\begin{table}[!ht]		
\begin{center}
\begin{tabular}{|c|c|c|c|c|c|c|c|}
\hline
\multicolumn{1}{|c}{Filter} & \multicolumn{1}{|c}{Pivot-WL} &
\multicolumn{1}{|c}{WFC (\%/yr)} & \multicolumn{1}{|c}{HRC (\%/yr)} & 
\multicolumn{1}{|c}{fit (\%/yr)} & 
\multicolumn{1}{|c}{mag/yr} & \multicolumn{1}{|c}{Ubeda} &
\multicolumn{1}{|c|}{2009.4}  \\

\hline \hline
F220W&  2257&   ...& -0.49&  0.66&  0.0072&  ...&   ...  \\
F250W&  2714&   ...& -0.68&  0.41&  0.0045&  ...&   ...  \\
F330W&  3362&   ...& -0.30&  0.22&  0.0024&  ...&   ...  \\
F344N&  3433&   ...&   ...&  0.20&  0.0022&  ...&   ...  \\
F435W&  4319& -0.07& -0.10&  0.15&  0.0016& 0.0027&   1.000\\
F475W&  4747& -0.05& -0.24&  0.16&  0.0017& 0.0034&   0.998\\
F502N&  5023&   ...&   ...&  0.17&  0.0018& 0.0012&   0.997\\
F555W&  5361& -0.26& -0.21&  0.18&  0.0020& 0.0018&   0.996\\
F550M&  5581& -0.20& -0.09&  0.19&  0.0020& 0.0024&   0.996\\
F606W&  5921& -0.15& -0.20&  0.20&  0.0021& 0.0011&   0.997\\
F625W&  6311& -0.23& -0.09&  0.20&  0.0021& 0.0030&   0.998\\
F658N&  6584&   ...&   ...&  0.19&  0.0021& 0.0039&   0.999\\
F660N&  6599&   ...&   ...&  0.19&  0.0021& 0.0039&   0.999\\
F775W&  7692& -0.06& -0.14&  0.12&  0.0014& 0.0016&   1.004\\
F814W&  8057& -0.06& -0.08&  0.10&  0.0010& 0.0019&   1.005\\
F892N&  8915&   ...&   ...&  0.06&  0.0007& ...&      1.007\\
F850LP& 9033& +0.06& -0.16&  0.06&  0.0007& 0.0014&   1.007\\
\hline
\end{tabular}
\end{center}
\caption{\textsl
{Sensitivity changes for each CCD filter. The units of the pivot wavelength
column are Angstroms. Columns 3--4 are the measured pre-SM4 loss rates in \%/yr
for WFC and HRC, respectively, while Column 5 is the average pre-SM4 loss rate
from the evaluation of the quartic fit to both cameras, as shown in
Figure~\ref{tchpre}. The negative signs are omitted in columns 5--7.
The narrow band measurements are generally unreliable and
are not tabulated, so that the fitted values must be used.
Columns 6 and 7 compare the fitted pre-SM4 loss rates in
$mag~yr^{-1}$ from the current analysis to the results from the 47 Tuc analysis
of \citet{ubeda13}. For the post-SM4 WFC observations, the final column
specifies the 2009.4 sensitivities relative to 2002.16 per the curve in
Figure~\ref{tchpost}.}} \label{table:losses} \end{table}

\begin{figure} 
\centering  
\includegraphics*[height=5in]{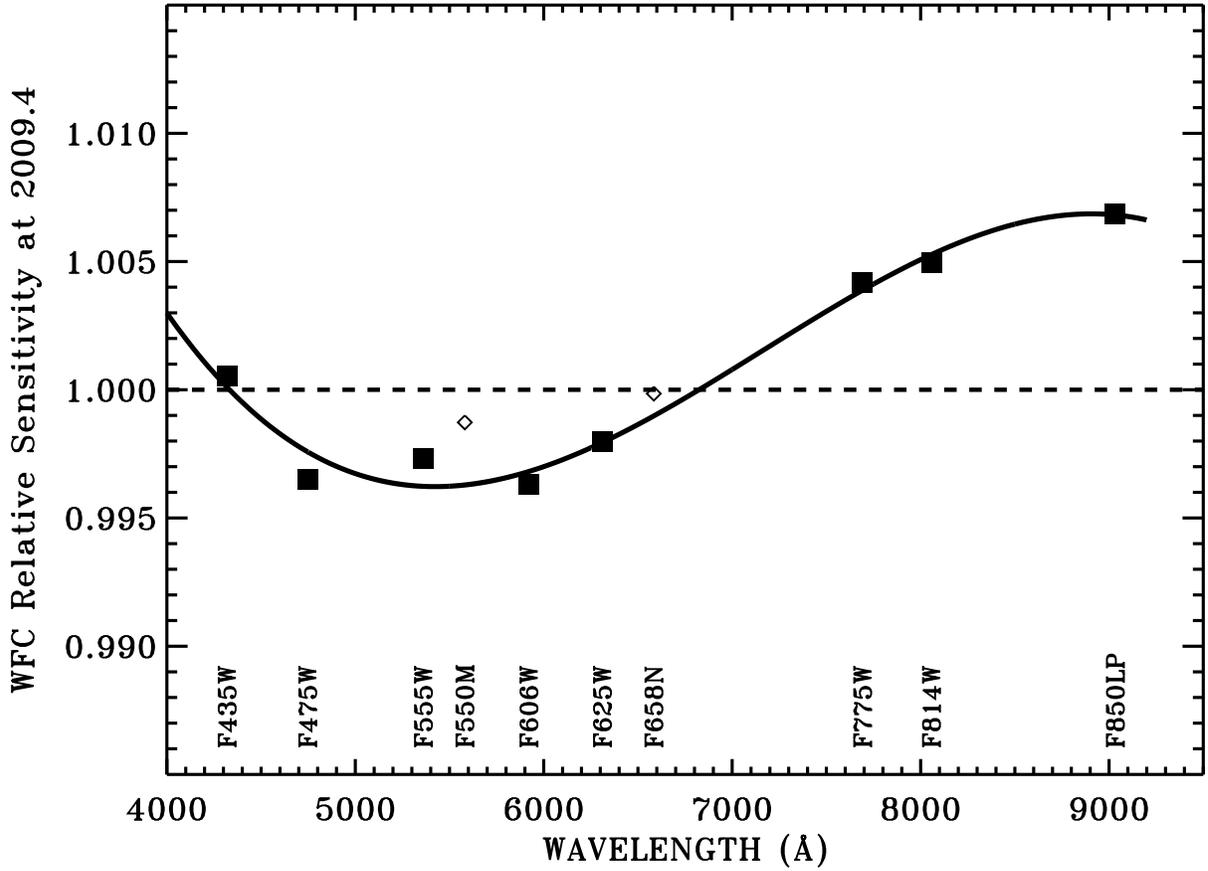}
\caption{\baselineskip=12pt
The WFC sensitivity correction at 2009.4 relative to the initial
sensitivity at 2002.16 as based on fitting the post-SM4 observations of
the three primary WD standard stars. The filled squares are the eight WFC
broadband filters, while the smooth curve is a cubic fit to these eight points.
The open diamonds are the medium and a narrow band filter.
\label{tchpost}} \end{figure}

The sensitivities for each WFC filter after SM4 are based on the fits to
$\sim$18 available WD observations for each of the eight broadband filters, as
illustrated by the solid lines in the Figure~\ref{tchfilt}.
The typical 1$\sigma$ rms scatter of 0.003 about the fits establishes the
uncertainty of an individual C/P value. These fits evaluated at 2009.4 appear in
Figure~\ref{tchpost}, where each value represents the sensitivity at 2009.4
relative to the initial sensitivity at 2002.16. However, the QE curve for the
-81C set point temperature is used for the post-SM4 calibration, so that the
Figure~\ref{tchpost} corrections must be applied to the post-2006.5 -81C QE, as
this QE is used to calculate the post-2009.4 sensitivities. Because this -81C QE
is designed to match the photometry to the -77C 2002.16 QE calibration, this
approach also matches the post-2009.4 fluxes to the 2002.16 flux scale. The
cubic fit in Figure~\ref{tchpost} has coefficients as in Equation 6 of a=1.1536,
b=-7.2839e-05, \mbox{c=1.0806e-08}, d=-5.0274e-13, and e=0 as a function of
wavelength in Angstroms, while the final column of Table~\ref{table:losses} has
the evaluations of this cubic at the pivot wavelengths. 

Originally, the post-SM4 calibration was set by adjusting
the post-SM4 gain to match the 2002.4 epoch photometry of 47 Tuc for F606W  
\citep{bohlinetal11}. Figure~\ref{tchpost} and Table~\ref{table:losses} show
that an additional small shift in sensitivity by 0.997 improves the match at
F606W, while the adjustments range from 0.996--1.007 for the other filters at
2009.4. The post-SM4 corrections at 2009.4 in Table~\ref{table:losses} along
with the average loss rate of 0.061\%/yr determine the post-SM4 $photflam$ flux
calibration for the time of the observation as detailed in Section 7.

\section{Bandpass Shifts}	

\begin{figure} 
\centering   
\includegraphics*[height=5in]{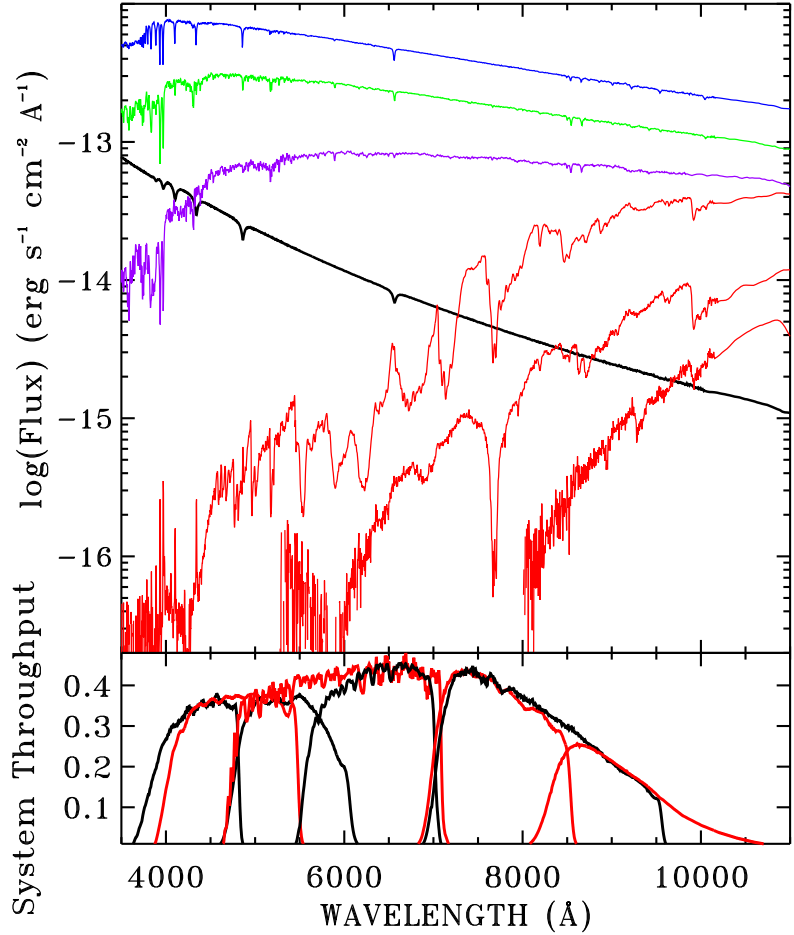}
\caption{\baselineskip=12pt
\textit{upper panel:} SEDs of stars used in this paper with the same color
coding as for Figure~\ref{eewfc}. The solid black line is the  WD GD153, and the
slopes of G191B2B and GD71 are similar. The red curves in order of bright to
faint are VB8, 2M0036+18, and 2M0559-14. For clarity, the SEDs for P330E
(green), KF06T2 (purple), 2M0036+18, and 2M0559-14 are scaled up by factors of
12, 8, 4, and 5, respectively. \textit{lower panel:} The system throughput of
the eight WFC broadband filters alternates between black and red lines.
\label{pltall}} \end{figure}

After fully correcting the photometry for the three primary WD standards, their
average C in relation to the synthetic predictions P, i.e. $\langle C/P\rangle$,
is unity after making the calibration updates in the following section. Any
systematic offset in the C/P for a cooler star is suggestive of errors in the
bandpass response function. For example, \citet{bohlin2012} found that the
bandpass function for F435W required a shift of the long wavelength cutoff of
+18~\AA\ at the \mbox{WFC1-1K} reference point; and there are lab measures for
all the broad filters that show bandpass variations of this order.
Figure~\ref{pltall} illustrates the SEDs of the stars used for bandpass
adjustments along with the system throughput of the WFC broadband filters.
Shifting the effective wavelength of a filter changes the relative response of
hot to cool stars. For example, moving F435W (shortest wavelength black line in
the lower panel)
to longer effective wavelengths decreases the response to hot WDs (black line
SED), while increasing the signal for the K star (purple line SED).

\begin{figure} 
\centering   
\includegraphics*[angle=90,height=5in]{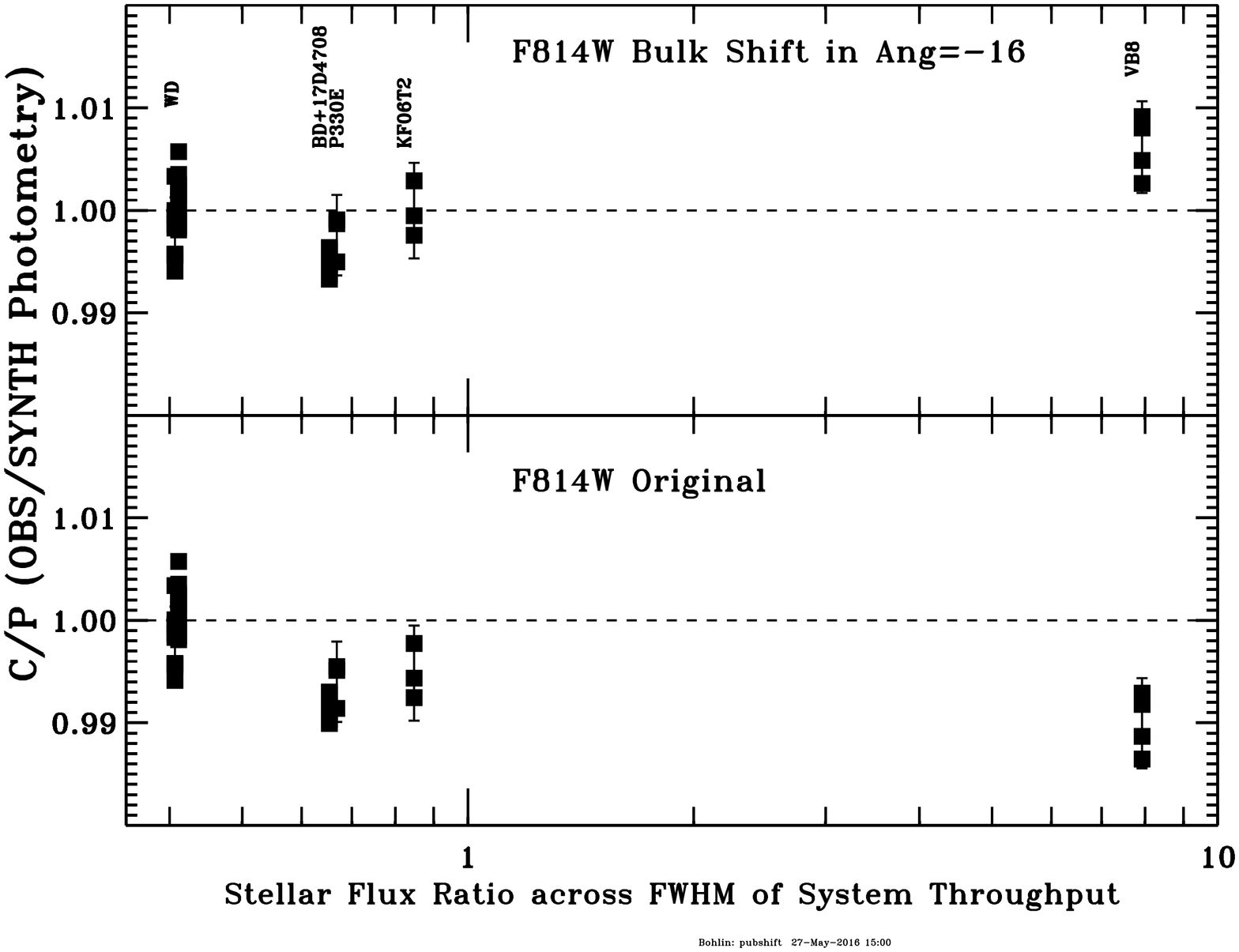}
\caption{\baselineskip=12pt
F814W C/P and 3$\sigma$ errors-in-the-mean at WFC1-1K
for the three primary WD standards and four cooler stars
shown as a function of the ratio of the stellar flux across the bandpass from
8920 to 7063~\AA. In the lower panel for the original bandpass function, the
cooler stars are systematically too faint in comparison to their synthetic
photometry P. In the top panel, a small shift of -16~\AA\ in the filter
transmission function reduces the cool star P values with respect to the WDs and
achieves agreement within 1\%, even for VB8. \label{shift}} \end{figure}

For both F435W and F814W, there are now at least 14 measurements of C/P spaced
over the WFC FOV for the K1.5III star KF06T2 with respect to the WD GD153
\citep{bohgrog}. Both stars are observed and differences are computed at each of
the field positions, so that the hot--cool star offsets are unaffected by any
flat fielding errors. The F814W C/P fractional offset of the K star from the WD
ranges up to $\sim$-1.5\%, while the K star is offset by -0.005 (-0.5\%) in C/P
at the WFC1-1K reference point based primarily on three observations of KF06T2.
Figure~\ref{shift} illustrates the -16~\AA\ bulk shift of the entire F814W
bandpass transmission function, which corrects this offset of -0.005 at WFC1-1K.
While the -16~\AA\ correction is precise for KF06T2, the C/P value for the G
star agrees within its $3\sigma$ error bar, the variable F star
BD+17$^{\circ}$4708 \citep{bohlinarlo15} is slightly low, and the less reliable
M star VB8 moves from 1\% low to $\sim$0.5\% high. Over the full FOV, the
maximum residual offset for KF06T2 with the new -16~\AA\ shift becomes -0.01,
i.e. a 1$\pm0.4\%$ error, which is now within our 1\% precision goal for the ACS
flux calibration for the first time. With a C/P based primarily on only three
F814W observations of the K star at WFC1-1K and without the confirming
observations spaced over the full FOV, an adjustment for a sub-percent bandpass
discrepancy is not justified. A bulk shift for F814W is better than shifting the
usually suspect long wave cutoff, because the photometry is relatively
insensitive to the cutoff wavelength; and an unreasonably large shift of $\sim$5
times the bulk shift of -16~\AA\ is required to make the same adjustment
of the K star photometry with respect to the WDs. The new bandpass reference
files for F435W and F814W are in the ACS calibration database.

\begin{figure} 
\centering   
\includegraphics*[angle=90,height=5in]{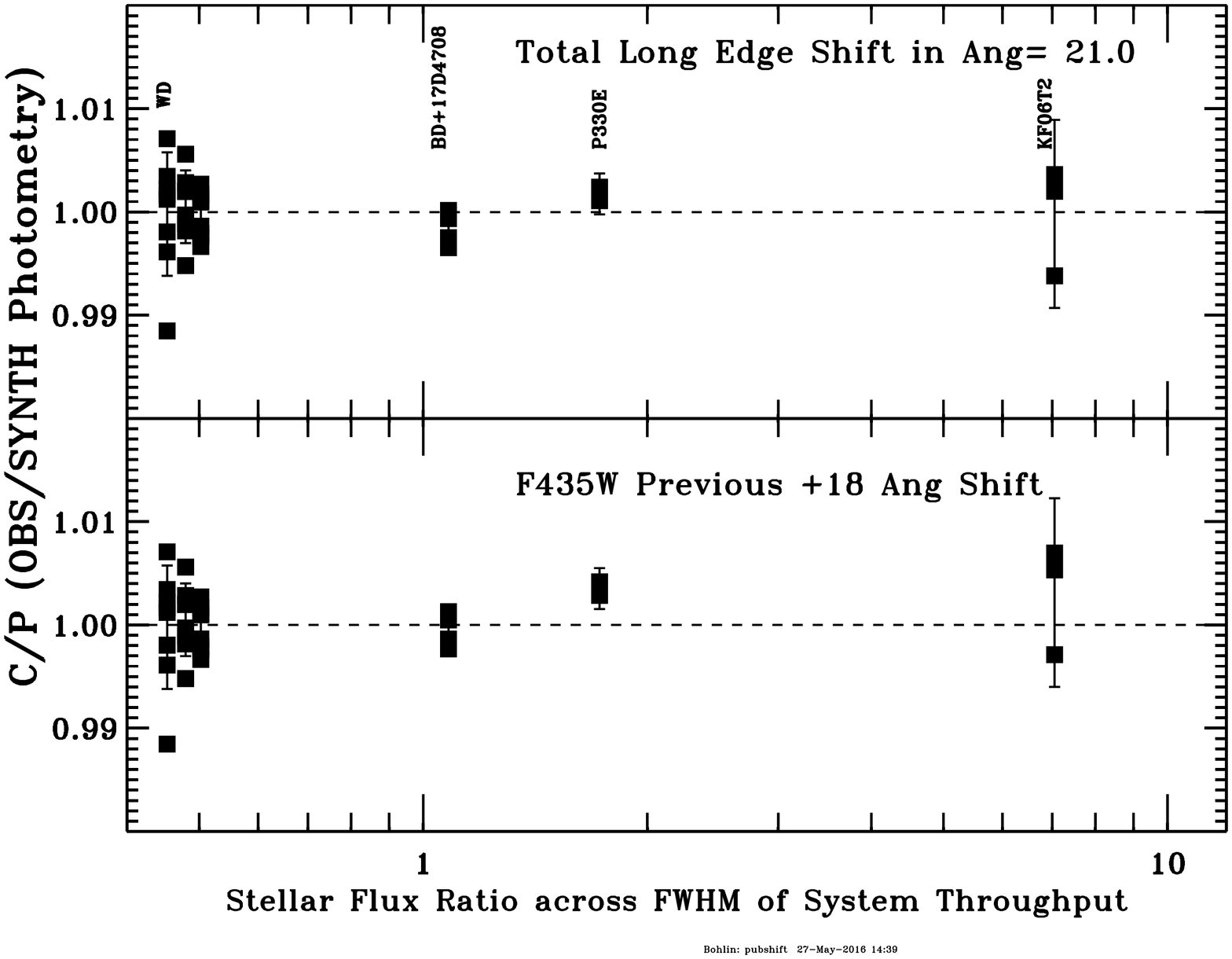}
\caption{\baselineskip=12pt
As in Figure~\ref{shift} for F435W for the three primary WD standards and three
cooler stars shown as a function of the ratio of the stellar flux across the
bandpass from 4815 to 3878~\AA. In the lower panel for the previous bandpass
function of \citet{bohlin2012}, the two coolest stars are systematically too
bright in comparison to their synthetic photometry P. In the top panel, a small
additional shift of +3~\AA\ in the long wavelength side of the filter
transmission function increases the cool star P values with respect to the WDs
and improves the hot/cool star agreement. \label{shift435}} \end{figure}

The new observations and newly reprocessed data for F435W leave a small residual
in hot vs. cool star C/P values at the WFC1-1K reference point; and a minor
adjustment is required from the +18~\AA\ long wavelength cutoff shift of
\citet{bohlin2012} to +21~\AA, as illustrated in Figure~\ref{shift435}. Even
though the maximum error is reduced, the F435W C/P offset
between KF06T2 and GD153 still has a 5\% range from -0.02 to +0.03 for 30
positions that are spaced around the FOV, which means that a bandpass for F435W
that varies over the FOV is required to achieve a 1\% photometric precision for
stars of different color temperature. For HRC, the lack of robust data sets
limit the validity of any bandpass adjustments.

\begin{figure} 
\centering   
\includegraphics*[height=5in]{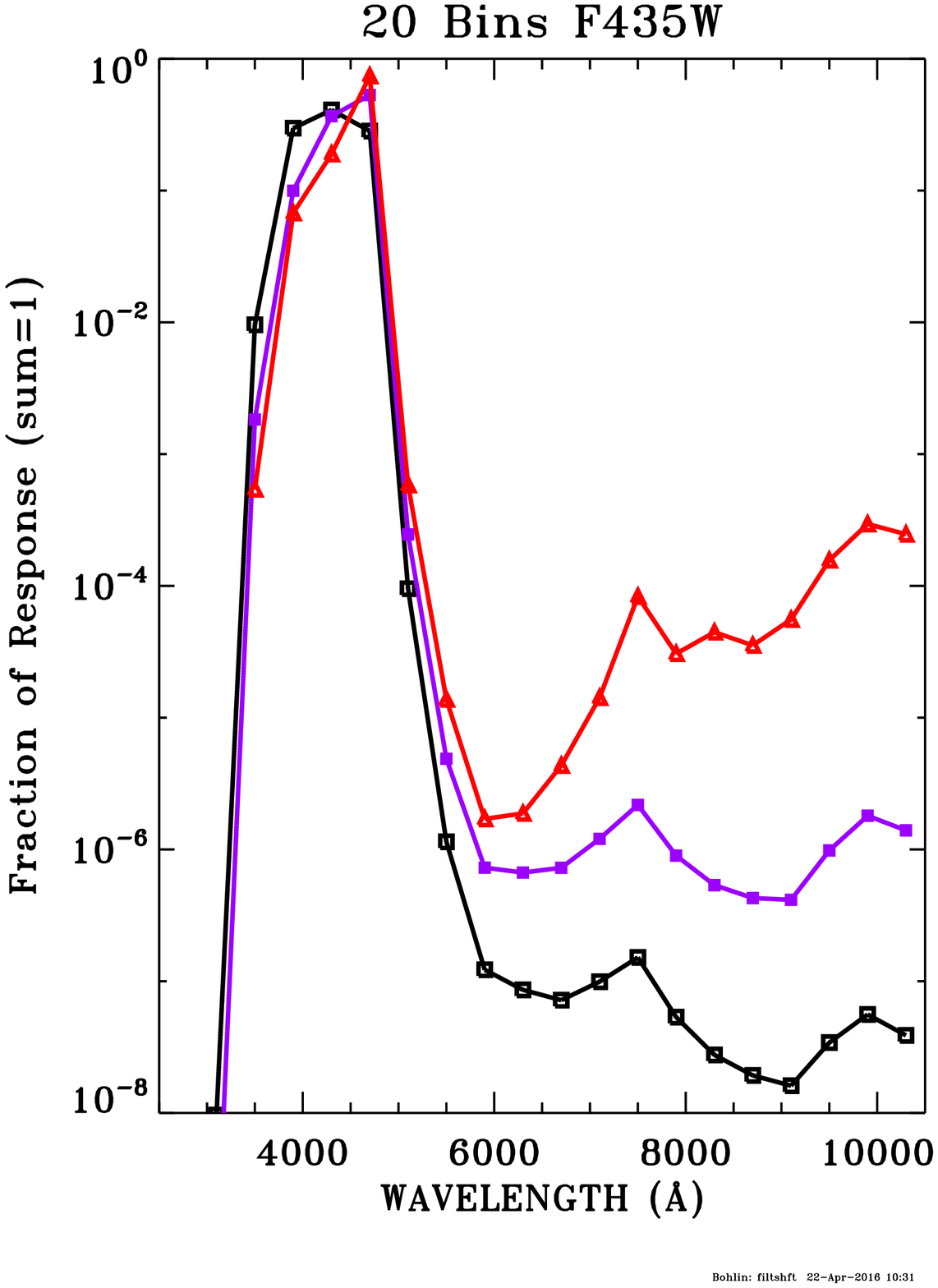}
\caption{\baselineskip=12pt
Fractional contribution to the WFC F435W signal in each of 20 bins spanning the
full range of the filter transmission. Black is for the hot WD G191B2B,
purple represents the K star KF06T2, and VB8 is red.
\label{filtshft}} \end{figure}

\subsection{Alternatives to Bandpass Shifts}	

In order to explain the discrepant comparison of ACS photometry to STIS fluxes
for cooler stars, \citet{bohlinisr07} considered and dismissed alternatives to
bandpass adjustments. In particular, out-of-band transmission errors are one
alternative; and Figure~\ref{filtshft} shows the contribution to the observed
ACS photometric signal as a function of wavelength for F435W in 20 equal
wavelength bins across the full measured range of the filter transmission. The
fractional contributions are computed from the CALSPEC SEDs for a hot WD
G191B2B, a K star KF06T2, and the M star VB8 using the ACS system throughput
$R$. The central three bins contain $>$99\% of the F435W signals for all three
stars, while the bin on the short wavelength shoulder of the response contains
most of the remaining contribution. As expected, the out-of-band red leak is
more than an order of magnitude higher for the K star than for the WD in the 12
points longward of $\sim$5900~\AA, while VB8 is another two orders of magnitude
larger. However, these 12 bins for KF06T2 are $\sim10^{-6}$ on average for a
total red leak fraction of $\sim10^{-5}$, which is a factor of 300 below the
0.3\% tweak achieved by updating the F435W filter edge shift from +18 to
+21~\AA. The ACS filter transmissions were all carefully measured in the lab
before launch and have exellent out-of-band rejections. \citet{bohlinisr07}
showed a similar figure for F625W; and all results for all eight of the
broadband filters are similar, with total red+blue leaks of $<10^{-4}$, i.e.
$<0.01\%$ for K and hotter stars. The transmission of F850LP was not measured
longward of the CCD QE cutoff of 11000~\AA.

\section{Absolute Flux Calibration}	

\subsection{Adjustments for Individual Filters}		

\begin{figure}   
\centering
\includegraphics*[angle=90,height=5.0in]{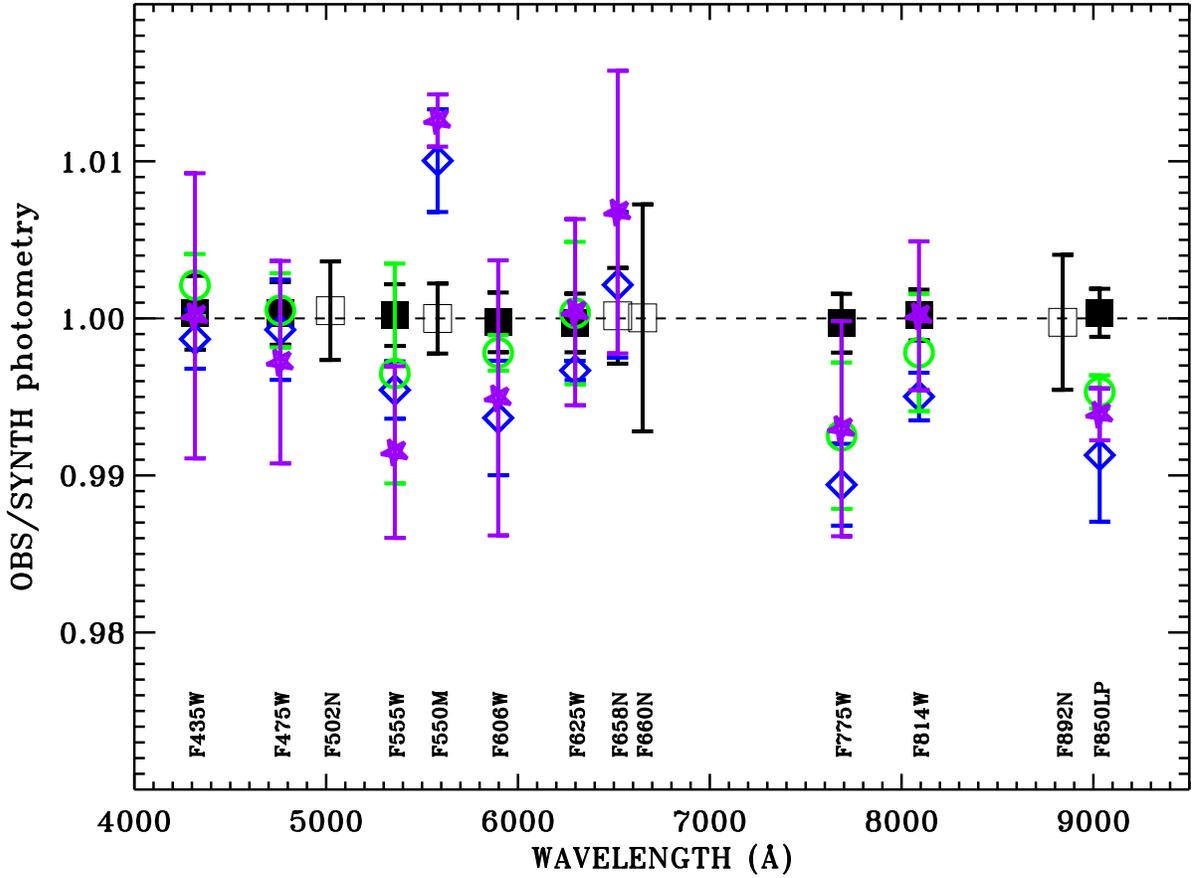}
\caption{\baselineskip=12pt
C/P count rate ratios vs. pivot wavelength for the stars observed with WFC after
making the corrections for the F435W and F814W filter shift and the corrections
in Table~\ref{table:resid} for the filter transmissions. The black squares
for the average of the three prime WDs are now all within 0.001 of unity, while
the averages and 3$\sigma$ errors-in-the-mean are shown for the F (blue diamond),
G (green circle), and K (purple star) standards. The less precise results for
the sparsely observed narrow and medium band filters are the open black squares.
\label{resid}} \end{figure}

Following any correction for filter bandpass shifts, the average residual
deviation of C/P from unity is corrected by multiplying each filter transmission
by the average residual in order to correct the P values that are computed from
the total system QE, i.e. $R$, which is the product of the throughput of all the
HST+ACS mirrors, the filter transmission, and the detector QE. The C values are
for the 1\arcsec\ (20 pixel) photometry from the \textit{*\_crj.fits} files, as
corrected for the pixel area maps (PAMs), the CTE losses, and the EE.
Table~\ref{table:resid} contains these average C/P residuals; and
Figure~\ref{resid} shows the final WFC average values (squares) for the three
primary WDs at unity with the cooler stars shown as colored symbols. There is no
consistent significant trend of the corrections with wavelength that would
mandate an update of the detector QE functions. The new filter throughputs
include the correction for these residuals, which are then propagated to the
final $photflam$ values according to Equation 3.

The fully corrected photometry
for the F, G, and K stars in the eight broadbands in Figure~\ref{resid} is
within the goal of 1\% agreement, except for one measure of the variable star
BD+17$^{\circ}$4708, which is consistently low in all eight broadbands. The
variable BD+17$^{\circ}$4708 \citep{bohlinarlo15} is  observed at different
epochs for the ACS numerator values and the STIS denominator. The sub-percent
internal consistency shown in  Figure~\ref{resid} is achieved for the first time
and is indicative of the consistency of the revised ACS flux calibration
relative to STIS (B14). 

\begin{deluxetable}{ccccc}	
\tablewidth{0pt}
\tablecolumns{5}
\tablecaption{Filter Transmission Updates}
\tablehead{
\multicolumn{1}{c}{~} &  \multicolumn{4}{c}{Residuals} \\
\colhead{Filter} &\colhead{WFC} &\colhead{3$\sigma$\tablenotemark{a}}  
&\colhead{HRC} 
&\colhead{3$\sigma$\tablenotemark{a}}}
\startdata
 F220W&   ...&    ...&  0.988&  0.005 \\
 F250W&   ...&    ...&  0.990&  0.005 \\
 F330W&   ...&    ...&  0.991&  0.003 \\
 F344N&   ...&    ...&  0.992&  0.005 \\
 F435W& 1.001&  0.002&  0.992&  0.003 \\
 F475W& 1.003&  0.002&  0.996&  0.004 \\
 F502N& 1.001&  0.003&  0.997&  0.007 \\
 F555W& 1.004&  0.002&  0.998&  0.001 \\
 F550M& 1.000&  0.002&  0.995&  0.005 \\
 F606W& 1.006&  0.002&  1.000&  0.002 \\
 F625W& 1.006&  0.002&  0.999&  0.004 \\
 F658N& 1.007&  0.003&  1.003&  0.009 \\
 F660N& 1.018&  0.007&  1.010&  0.008 \\
 F775W& 1.008&  0.002&  1.001&  0.001 \\
 F814W& 0.994&  0.002&  1.000&  0.003 \\
 F892N& 1.006&  0.004&  1.015&  0.009 \\
F850LP& 1.008&  0.002&  1.007&  0.005 \\
\enddata
\tablenotetext{(a)}{The ~3$\sigma$ uncertainties of the residuals are 
errors-in-the-mean.}
\label{table:resid}
\end{deluxetable}

\subsection{Uncertainty in the Flux Calibration}		

\subsubsection{Uncertainty at WFC1-1K}

The ACS absolute flux calibrations $S$ are computed with the system QE, i.e.
$R$, according to Equation 3, where the $R$ functions are determined from C/P,
as  in Section 7.1. Because the infinite aperture count rate C is computed from
the 20 pixel aperture photometry that has the best repeatability as
$C=N_{20}/E_{20}$, the uncertainty in $S$ is the uncertainty in the average
$N_{20}/E_{20}$, where the repeatability of $N_{20}$ from Table~\ref{table:sig}
for the broadband WFC filters is 0.30--0.53\% for the 20 pixel radius photometry.
The uncertainty in the average $N_{20}$ rms scatter is reduced by the typical
number of 23 broadband observations of the primary WDs to a maximum of
$0.53/\sqrt{23}=0.11~\%$, while the uncertainty for $E_{20}$ from Section 4.3 is
0.1\%. Thus, the formal uncertainties in the ACS broadband $photflam$ values that
are based on the 20 pixel radius photometry are all $\leq$0.15\%. For the four
narrow band filters, the statistical measurement uncertainties are somewhat
larger according to their error bars in Figure~\ref{resid}.

However, this systematic uncertainty increases, if smaller aperture photometry
is required. For example, the maximum of 2.23\% from Table~\ref{table:sig} for
F775W combined with the EE uncertainty of 0.4\% makes a total WFC three pixel
flux calibration systematic error of $\sqrt{{2.23^2}/23+0.4^2}=0.6\%$, which has
a floor value of 0.4\%, even if the ACS observational data set were greatly
increased. The EE uncertainty is not likely to benefit significantly from more
observations, as that 0.4\% is already based on a smooth fit to a large
data set.

The above error estimates are relative to the STIS flux calibration that
determines the $P$ values in C/P, and the external STIS uncertainties are
presented in figure 14 of B14. Thus, the systematic ACS and
STIS errors must be combined to get the full systematic uncertainty. For
example, the STIS uncertainty at the 9033~\AA\ pivot wavelength of F850LP is
0.4\% relative to the reference wavelength of 5556~\AA. Combining the STIS 0.4\%
with the ACS WFC uncertainties of 0.87/$\sqrt{23}$ from Table~\ref{table:sig}
and 0.2\% in EE for a five pixel radius makes a total of $\pm$0.5\% for F850LP
relative to 5556~\AA. A grand total absolute uncertainty should also include the
0.5\% uncertainty in the reference flux for Vega at 5556~\AA\ \citep{bohlin14}.
A proper  formal STIS error analysis should utilize the covariant matrix
$WDcovar.fits$ from B14 that is available in the CALSPEC
archive.

\subsubsection{Uncertainty around the Full Field of View}

If the flat fields were perfect for stars of any spectral type. then the above
discussion of uncertainties at the WFC1-1K reference point would apply
everywhere. However, there are sometimes problems with flat field uniformity, as
discussed in \citet{bohgrog} and in Sections 4.5 and 6 above. While the bandpass
adustments of Section 6 brings the uniformity to better than 1\% for F814W,
F435W shows larger variations of -2 to +3\% in GD153-KF06T2 differential
photometry. The throughput variations around the F435W FOV for the individual
stars are -4.2 to +1.7\% for the WD and -2.7 to +0.9\% for the K star. The
existing flat fields are all based on the average stellar type in 47 Tuc
\citep{mack02}; and the smaller error range for the K star KF06T2 suggests that
the average type of 47 Tuc is closer to K than to the 40,000~K WD GD153. In the
presence of flat fielding errors, the absolute flux calibrations provided here
for the hot WDs at the WFC1-1K reference point may differ somewhat from an
average over the full field; but there is now at least one point where the
calibration is fully characterized. If the -4.2 to +1.7\% WD error range for
F435W is a worst case, then -1.2\% would be the largest systematic error of the
WFC1-1K calibraton with respect to the FOV average. Except for F435W and F814W,
there is little data on flat fielding errors for stars that differ from the mean
color of the 47 Tuc stars that define the flat fields. However, 
\citet{bohlin2012} demonstrated that the meager archival photometry for F775W
varied by $<$0.5\% for the WD G191B2B at seven different positions in the FOV.

\subsection{QE Reference Files}		

The changing ACS sensitivity is incorporated in the pipeline data processing via
the Synphot QE reference files for the WFC and HRC CCD detectors, which are
structured with wavelength in column 1, a default QE in column 2 for unspecified
dates, and then pairs of QE columns, where the actual QE at any date is linearly
interpolated in time between the dates for columns 3 and 4, columns 5--6, etc.
The $photflam$ calibration constants $S$ in the ACS data headers are computed
for the time of the observation using the new QE reference files
acs\_hrc\_ccd\_mjd\_016\_syn.fits or the pair of identical WFC files:
acs\_wfc\_ccd1\_mjd\_022\_syn.fits and acs\_wfc\_ccd2\_mjd\_022\_syn.fits. 

This calibration scheme is reflected in Table~\ref{table:photflam}, where the
default column 2 in the reference file is omitted; and the \textit{photflam}
values are those calculated using the QE vectors for the dates of
the columns in the QE reference files.
In the case of HRC, there is only one pair of QE columns with dates of
2002.16 and the end of HRC operations at 2007.1 (MJD=54136). 

For WFC, the QE changed at 2006.5 with the switch to the side 2 electronics
chain when the operating temperature was lowered, so that the first pair of QE
columns for -77C cover the date range 2002.16--2006.5 (MJD 52334--53919), while
the second pair for -81C is for 2006.5--2009.4 (MJD 53920--54976); and the third
pair covers 2009.4--2021.0 (MJD 54977--59214). The loss rates for the first two
pairs of WFC columns through 2009.4 and for HRC are those specified in 
Table~\ref{table:losses}, while the difference between the WFC columns for
2009.4--2021.0 represents
the average post-SM4 loss rate of 0.061\%/yr.
The ACS CCD channels were
inactive from early 2007 until after the SM4 servicing mission at 2009.4, i.e.
MJD 54977; but defining a correction during this dead
time causes no harm. 

The QE of the HRC detector at 2002.16 in the reference file was updated by
\citet{bohlin2012}, but the WFC QE was last adjusted by \citet{bohlinisr07}.
Thus, the changes from the previous calibration for the 2002.16
\textit{photflam} values in Table~\ref{table:photflam} are due only to
the filter residuals of Table~\ref{table:resid} and to the two
WFC bandpass adjustments.
The differences between columns 3 and 4 in the Table~\ref{table:photflam}
$photflam$ values represent the change from the set point temperature of -77C to
-81C at 2006.5 \citep{mack07}, while the differences from column 2 to column 6
at 2009.4 include both the \citet{mack07} adjustment for the colder temperature
and the adjustments in final column of Table~\ref{table:losses}. The second
value of each pair represents the increased \textit{photflam} that compensates for the
loss in sensitivity from the date of the first \textit{photflam} of the pair.

The infinite aperture $photflam$ calibrations of Table~\ref{table:photflam} are
included in the ACS data headers but must be converted for
practical usage of smaller aperture photometry via Equation 1, where $\langle
F\rangle={S~N_e}=S~N_i/E_i$. N$_i$ is photometry for the $ith$ aperture radius
and $E_i$ is the encircled energy EE correction from Table~\ref{table:eewfc} or
Table~\ref{table:eehrc}. In case such a table for any future revision is not at
hand or to avoid the necessity of consulting the literature to calibrate
aperture photometry, the $photflam(i)=S/E_i$ or just the $E_i$ should also be
included in the $fits$ data headers for the i=3, 5, 10, and 20 aperture radii.

\subsection{Implications of Updated Calibration}		

New data contributes to improved mean EE values, which are now provided for
aperture radii as small as one pixel. The robust set of new post-SM4 data
reveals a sensitivity loss for all filters of 0.061\%/yr, so that by 2017, the
change in sensitivity will reach 0.5\% from the time of the 2009 SM4.  Another
new body of data for F435W and F814W demonstrates the need for small bandpass
adjustments for those two filters.

Column 2 for WFC or column 8 for HRC in Table~\ref{table:photflam} can be
compared to the $photflam$ results of \citet{sirianni05} for the first years of
ACS operations. The current WFC values range from 5\% smaller for F660N to 1\%
larger for F550M, while the eight broadband filters are 1--3\% smaller. For HRC,
differences range from 0--3\% smaller, except for F344N, where the new
calibration is 3\% larger.
These differences are due mostly to changes in the CALSPEC reference SEDs and to
a refined and expanded set of ACS observations of these standard stars.
\citet{sirianni05} utilized many observations of GRW+70$^{\circ}$5824, which has
a poor, noisy CALSPEC SED that can affect the synthetic photometry, especially
for narrow bandpasses. Observations of GRW+70$^{\circ}$5824 are not used for the
current, new $photflam$ results in Table~\ref{table:photflam}.


\begin{deluxetable}{cccccccccc}	
\rotate
\tablewidth{0pt}
\tablecolumns{10}
\tablecaption{ACS infinite aperture calibration constants $photflam$, 
i.e.~$S$ in units of $erg~s^{-1}~cm^{-2}~\AA^{-1}/(e~s^{-1})$. 
The applicable
dates are listed as MJD and as fractional year, along with the WFC CCD set point
temperature. The $S$ at any date is
the linear interpolation between pairs of columns.}
\tablehead{
\multicolumn{1}{c}{~} & \multicolumn{7}{c}{WFC} & \multicolumn{2}{c}{HRC}  \\
\multicolumn{1}{c}{Filter} & \multicolumn{1}{c}{MJD52334}  &
\multicolumn{1}{c}{MJD53919}  & \multicolumn{1}{c}{MJD53920}  & 
\multicolumn{1}{c}{MJD54976}  & \multicolumn{1}{c}{MJD54977}  &
\multicolumn{1}{c}{MJD59214}  & \multicolumn{1}{c}{~}         &
\multicolumn{1}{c}{MJD52334}  & \multicolumn{1}{c}{MJD54136} \\

\multicolumn{1}{c}{~} & \multicolumn{1}{c}{2002.16}  &
\multicolumn{1}{c}{2006.5}  & \multicolumn{1}{c}{2006.5}  & 
\multicolumn{1}{c}{2009.4}  & \multicolumn{1}{c}{2009.4}  &
\multicolumn{1}{c}{2021.0} &  \multicolumn{1}{c}{~}       &
\multicolumn{1}{c}{2002.16}  & \multicolumn{1}{c}{2007.1} \\

\multicolumn{1}{c}{~} & \multicolumn{1}{c}{-77C}  &
\multicolumn{1}{c}{-77C}  & \multicolumn{1}{c}{-81C}  & 
\multicolumn{1}{c}{-81C}  & \multicolumn{1}{c}{-81C}  & 
\multicolumn{1}{c}{-81C}  & \multicolumn{1}{c}{~}     &
\multicolumn{1}{c}{~}  & \multicolumn{1}{c}{~}}
\startdata
 F220W&          &  &  &  &  &  &					    & 8.063e-18& 8.333e-18  \\
 F250W&          &  &  &  &  &  &					    & 4.636e-18& 4.731e-18  \\
 F330W&          &  &  &  &  &  &					    & 2.210e-18& 2.235e-18  \\
 F344N&          &  &  &  &  &  &					    & 2.200e-17& 2.222e-17  \\
 F435W& 3.062e-19& 3.082e-19& 3.155e-19& 3.169e-19&  3.135e-19& 3.157e-19 & & 5.318e-19& 5.358e-19  \\
 F475W& 1.781e-19& 1.793e-19& 1.828e-19& 1.837e-19&  1.819e-19& 1.832e-19 & & 2.884e-19& 2.907e-19  \\
 F502N& 5.131e-18& 5.168e-18& 5.257e-18& 5.283e-18&  5.237e-18& 5.274e-18 & & 7.977e-18& 8.043e-18  \\
 F550M& 3.906e-19& 3.939e-19& 3.991e-19& 4.013e-19&  3.973e-19& 4.002e-19 & & 5.797e-19& 5.851e-19  \\
 F555W& 1.920e-19& 1.935e-19& 1.964e-19& 1.974e-19&  1.955e-19& 1.969e-19 & & 2.985e-19& 3.012e-19  \\
 F606W& 7.666e-20& 7.728e-20& 7.824e-20& 7.867e-20&  7.778e-20& 7.834e-20 & & 1.258e-19& 1.270e-19  \\  
 F625W& 1.168e-19& 1.178e-19& 1.191e-19& 1.197e-19&  1.183e-19& 1.191e-19 & & 1.942e-19& 1.960e-19  \\
 F658N& 1.940e-18& 1.956e-18& 1.976e-18& 1.988e-18&  1.962e-18& 1.976e-18 & & 3.320e-18& 3.352e-18  \\
 F660N& 5.077e-18& 5.120e-18& 5.172e-18& 5.201e-18&  5.134e-18& 5.171e-18 & & 8.754e-18& 8.838e-18  \\
 F775W& 9.826e-20& 9.878e-20& 1.000e-19& 1.004e-19&  9.912e-20& 9.983e-20 & & 1.926e-19& 1.938e-19  \\
 F814W& 6.943e-20& 6.974e-20& 7.082e-20& 7.103e-20&  7.017e-20& 7.067e-20 & & 1.268e-19& 1.274e-19  \\
 F850LP&1.489e-19& 1.493e-19& 1.528e-19& 1.531e-19&  1.514e-19& 1.525e-19 & & 2.247e-19& 2.254e-19  \\
 F892N& 1.475e-18& 1.479e-18& 1.510e-18& 1.512e-18&  1.496e-18& 1.506e-18 & & 2.436e-18& 2.443e-18  \\
\enddata
\label{table:photflam}
\end{deluxetable}
 
\section{Conclusions and Recommendations}

For stellar types of K and hotter, Figure~\ref{resid} demonstrates the
achievement of the goal of 1\% precision of the ACS/WFC flux calibration
relative to the STIS standard stars for the eight well observed broadband
filters. Small adjustments of the filter transmission functions for F555W,
F775W, and F850LP similar to those for F435W and F814W would bring the agreement
for all eight filters within 0.5\% at the WFC1-1K reference point; but before
making those three new transmission corrections, confirming observations should
be obtained like those mapping the FOV for F435W and F814W.

If precision photometry of the coolest stars of type M and later is important,
then the WFC calibration for these types could be improved by more observations
both with ACS/WFC and STIS. Action to augment the header keywords, to solve the 
drizzle problem, and to derive color dependent F435W flat fields is not yet
scheduled. However, the new \textit{photflam} flux calibration should be
implemented by the time this article is published. The \textit{photflam} data
header keyword for the infinite aperture includes the changes with time of
Section 5, which are encapsulated in the reference files
acs\_hrc\_ccd\_mjd\_016\_syn.fits and acs\_wfc\_ccd*\_mjd\_022\_syn.fits, where
the asterisk can be either of the identical 1 or 2 files. The revised filter
throughput curves are named by filter and detector and are also available as
reference files, while EE values must be read from Tables~\ref{table:eewfc} and
Table~\ref{table:eehrc} herein.

ACKNOWLEDGEMENTS

Thanks to J. Mack for re-drizzling the jcr601081\_drz.fits image and for
suggesting that the WHT drizzle extent can be used to distinguish between good
and bad drz photometry. Wayne Landsman provided the IDL photometry routine
\textit{apphot.pro}. This research has made use of the SIMBAD database,
operated at CDS, Strasbourg, France.

\appendix
\section{Drizzle $drz$ Photometry}

In principle, extracting photometry from the geometrically corrected drizzled 
\textit{*\_drz.fits} files is more straightforward than using the
\textit{*\_crj.fits} files. There is no need to apply PAMs and the plate scale
is constant at exactly 0.05\arcsec\ per pixel for WFC (0.025\arcsec\ for HRC).
However, the \textit{*\_drz.fits} files are problematic when combining
pairs of images; and use of the default pipeline
products requires extreme caution. The second extention of the \textit{drz} product is the weight (WHT)
image, which offers clues to the drizzle validity. If the pair of
\textit{*\_flt.fits} images that are combined into the \textit{drz} have
slightly different or misaligned PSFs or even mis-matched backgrounds
\citep{avila15}, a reduced WHT coincident with the core of the star usually
indicates improper cosmic-ray flagging. A data quality (DQ) parameter can be
defined as the percent drop from maximum to minimum WHT weight in the 3x3 set of
pixels centered on the star. Setting a DQ=13.1\% limit provides a fairly good
dividing line between good and bad \textit{drz} photometry for pairs of images,
and most of the errors are flagged without flagging many of the valid
\textit{drz} results. The \textit{crj} photometry is validated by the small
$\sigma_{20}<$0.5\% in Table~\ref{table:sig} for the baseline 20 pixel
photometry.

\begin{figure}   
\centering
\includegraphics*[angle=90,height=5.0in]{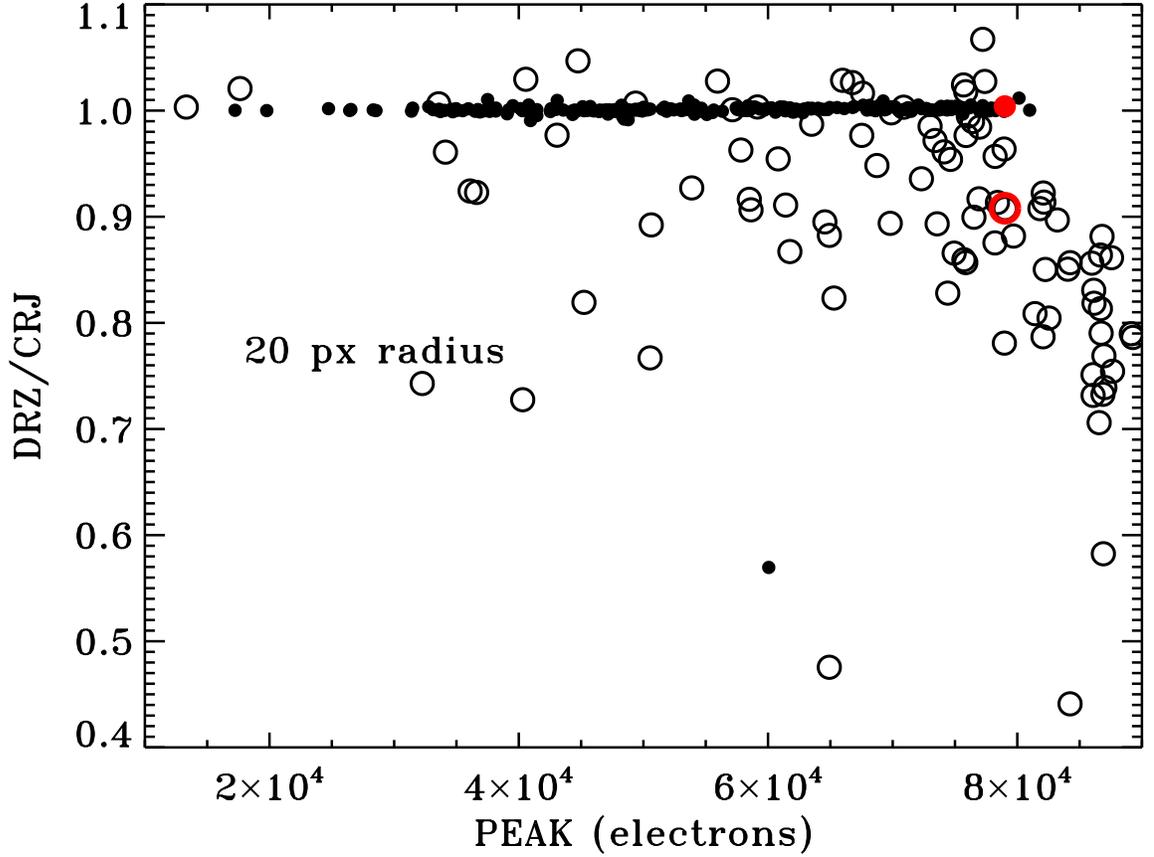}
\caption{\baselineskip=12pt
Comparison of 20 pixel photometry from \textit{crj} and \textit{drz} files 
for the WFC standard
star observations. The 239 filled black circles are the ratios for 
\textit{drz} photometry with good weight (WHT) arrays, while the open black
circles represent the same ratio but with a WHT DQ parameter of 
greater than 0.131. The erroneous \textit{drz} results become more frequent
above a peak signal level of about half of the CCD saturation level of 83,000
electrons. The open and filled red circles represent pipeline processing and a
special reprocessing with relaxed drizzle constrains, respectively,
which suggests that the default pipeline drizzle parameters could be improved.
\label{crjdrz}} \end{figure}

Figure~\ref{crjdrz}, compares the WFC \textit{drz} with the baseline 
\textit{crj} 1\arcsec\ photometry of the standard stars as a function of the
response in the peak pixel of each observation. There are 239 unflagged
comparisons shown as the filled circles that are all within 1\% of perfect
agreement. However, the 95 flagged ratios are 28\% of the total, are usually in
error by more than 1\%, and occassionally are in error by more than a factor of
two. The \textit{crj} pipeline product from the
MAST data archives is not always perfect, even for the
short standard star exposures, where hot pixels are not a problem. Of the 334
exposures in Figure~\ref{crjdrz}, one (\textit{jcga02081\_crj.fits}) requires
special re-processing with relaxed cosmic-ray (CR) rejection parameters, because
the CR-split pair is slightly misaligned.


One case of combining the jcr601lhq and jcr601liq \textit{flt} files  to make
the jcr601081 \textit{crj} and \textit{drz} files is investigated in detail. The
pipeline \textit{drz/crj} photometry ratio is 0.91, while the peak is 79000
(open red circle in  Figure~\ref{crjdrz}). Reprocessing this observation with
relaxed drizzle constraints for cosmic-ray rejection of driz\_cr\_scale=(1.5,
1.2), instead of the pipeline default (1.2, 0.7) values \citep{gonzaga12}
produces a revised ratio of unity (filled red circle in Figure~\ref{crjdrz}).
The ACS drizzle product could be much more useful to the science community by a
relaxation of the pipeline parameters that would bring \textit{drz}
photometry into agreement with the robust \textit{crj} results.

The ACS/CCD flux calibrations at the WFC1-1K reference point for the good,
unflagged WFC \textit{drz} photometry are equivalent to the \textit{crj}
calibrations, except for the small, three pixel radius, where the \textit{drz}
photometry and EE are systematically $\sim$0.5\% smaller. The average 1\arcsec
~(20 pixel) \textit{drz} photometry for the three WDs is the same as the
\textit{crj} results within 0.1\%; and the \textit{drz} flux calibration is the
same as the \textit{crj} calibration in Table~\ref{table:resid} to 0.1\%. Except
for the 0.5\% difference for the  three pixel radius, any systematic difference
between \textit{drz} and \textit{crj} WFC photometry at the WFC1-1K reference
point is less than the uncertainty; and the \textit{crj} calibration is accurate
for \textit{drz} results. At other locations in the FOV, the sparse data set
combined with the problematic \textit{drz} results prohibit reliable measures of
any differences between  \textit{drz} and \textit{crj} EE.

\section{Encircled Energies}

Table~\ref{table:eewfc} and Table~\ref{table:eehrc} include EE values for 12
aperture radii as computed according to the quartic fitting procedure of Section
4.3.

\newpage

\bibliographystyle{apj}
\bibliography{../../../pub/paper-bibliog}

\begin{thebibliography}{}
\expandafter\ifx\csname natexlab\endcsname\relax\def\natexlab#1{#1}\fi

\bibitem[{{Anderson} \& {Bedin}(2010)}]{Anderson2010}
{Anderson}, J., \& {Bedin}, L.~R. 2010, \pasp, 122, 1035

\bibitem[{{Avila} {et~al.}(2015){Avila}, {Hack}, {Cara}, {Borncamp}, {Mack},
  {Smith}, \& {Ubeda}}]{avila15}
{Avila}, R.~J., {Hack}, W., {Cara}, M., {et~al.} 2015, in Astronomical Society
  of the Pacific Conference Series, Vol. 495, Astronomical Data Analysis
  Software an Systems XXIV (ADASS XXIV), ed. A.~R. {Taylor} \& E.~{Rosolowsky},
  281

\bibitem[{{Bohlin}(2007)}]{bohlinisr07}
{Bohlin}, R.~C. 2007, {Photometric Calibration of the ACS CCD Cameras,
  Instrument Science Report, ACS 2007--06, (Baltimore: STScI)}, Tech. rep.

\bibitem[{{Bohlin}(2011)}]{bohlin2011}
---. 2011, {Flux Calibration of the ACS CCD Cameras II. Encircled Energy
  Correction, Instrument Science Report, ACS 2011--02, (Baltimore: STScI)},
  Tech. rep.

\bibitem[{{Bohlin}(2012)}]{bohlin2012}
---. 2012, {Flux Calibration of the ACS CCD Cameras IV. Absolute Fluxes,
  Instrument Science Report, ACS 2012--01, (Baltimore: STScI)}, Tech. rep.

\bibitem[{{Bohlin}(2014)}]{bohlin14}
---. 2014, \aj, 147, 127

\bibitem[{{Bohlin} \& {Anderson}(2011)}]{bohlinjay2011}
{Bohlin}, R.~C., \& {Anderson}, J. 2011, {Flux Calibration of the ACS CCD
  Cameras I. CTE Correction, Instrument Science Report, ACS 2011--01,
  (Baltimore: STScI)}, Tech. rep.

\bibitem[{{Bohlin} {et~al.}(2014){Bohlin}, {Gordon}, \&
  {Tremblay}}]{bohlinetal14}
{Bohlin}, R.~C., {Gordon}, K.~D., \& {Tremblay}, P.-E. 2014, \pasp, 126, 711
  (B14)

\bibitem[{{Bohlin} \& {Grogin}(2015)}]{bohgrog}
{Bohlin}, R.~C., \& {Grogin}, N. 2015, {Flat Field Determinations using an
  Isolated Point Source, Instrument Science Report, ACS 2015--07, (Baltimore:
  STScI)}, Tech. rep.

\bibitem[{{Bohlin} \& {Landolt}(2015)}]{bohlinarlo15}
{Bohlin}, R.~C., \& {Landolt}, A.~U. 2015, \aj, 149, 122

\bibitem[{{Bohlin} {et~al.}(2011){Bohlin}, {Mack}, \& {Ubeda}}]{bohlinetal11}
{Bohlin}, R.~C., {Mack}, J., \& {Ubeda}, L. 2011, {Flux Calibration of the ACS
  CCD Cameras III. Sensitivity Changes over Time, Instrument Science Report,
  ACS 2011--03, (Baltimore: STScI)}, Tech. rep.

\bibitem[{{Bohlin} \& {Proffitt}(2015)}]{bohlin15}
{Bohlin}, R.~C., \& {Proffitt}, C.~R. 2015, {Improved Photometry for G750L,
  Instrument Science Report, STIS 2015--01, (Baltimore: STScI)}, Tech. rep.

\bibitem[{{Gilliland}(2004)}]{gillil2004}
{Gilliland}, R.~L. 2004, {ACS CCD Gains, Full Well Depths, and Linearity up to
  and Beyond Saturation, Instrument Science Report ACS 2004--01, (Baltimore:
  STScI)}, Tech. rep.

\bibitem[{{Gilliland} {et~al.}(2006){Gilliland}, {Bohlin}, \&
  {Mack}}]{gilliland06}
{Gilliland}, R.~L., {Bohlin}, R., \& {Mack}, J. 2006, {WFC L-Flats Post
  Cooldown, Instrument Science Report, ACS 2006--06, (Baltimore: STScI)}, Tech.
  rep.

\bibitem[{{Gonzaga} {et~al.}(2012){Gonzaga}, {Hack}, {Fruchter}, \&
  {Mack}}]{gonzaga12}
{Gonzaga}, S., {Hack}, W., {Fruchter}, A., \& {Mack}, J. 2012, {The DrizzlePac
  Handbook, (Baltimore:STScI)}

\bibitem[{{Grogin} {et~al.}(2010){Grogin}, {Lim}, {Maybhate}, {Hook}, \&
  {Loose}}]{grogin10}
{Grogin}, N.~A., {Lim}, P.~L., {Maybhate}, A., {Hook}, R.~N., \& {Loose}, M.
  2010, in Hubble after SM4. Preparing JWST, Post-SM4 ACS/WFC Bias Striping:
  Characterization And Mitigation, 2010 STScI Calibration Workshop, eds. S.
  Deustua and C. Oliveira (Baltimore: STScI), 481

\bibitem[{{Krist} \& {Hook}(2004)}]{krist04}
{Krist}, J., \& {Hook}, R. 2004, {The Tiny Tim User's Guide, Version 6.3
  (Baltimore: STScI) http://www.stsci.edu/software/tinytim, p. 8}, Tech. rep.

\bibitem[{{Mack} {et~al.}(2002){Mack}, {Bohlin}, {Gilliland}, {van der Marel},
  {Blakeslee}, \& {de Marchi}}]{mack02}
{Mack}, J., {Bohlin}, R., {Gilliland}, R., {et~al.} 2002, {ACS L-Flats for the
  WFC, Instrument Science Report, ACS 2002--08, (Baltimore:STScI)}, Tech. rep.

\bibitem[{{Mack} {et~al.}(2007){Mack}, {Gilliland}, {Anderson}, \&
  {Sirianni}}]{mack07}
{Mack}, J., {Gilliland}, R.~L., {Anderson}, J., \& {Sirianni}, M. 2007, {WFC
  Zeropoints at -80C, Instrument Science Report, ACS 2007--02,
  (Baltimore:STScI)}, Tech. rep.

\bibitem[{{Scolnic} {et~al.}(2014){Scolnic}, {Rest}, {Riess}, {Huber}, {Foley},
  {Brout}, {Chornock}, {Narayan}, {Tonry}, {Berger}, {Soderberg}, {Stubbs},
  {Kirshner}, {Rodney}, {Smartt}, {Schlafly}, {Botticella}, {Challis},
  {Czekala}, {Drout}, {Hudson}, {Kotak}, {Leibler}, {Lunnan}, {Marion},
  {McCrum}, {Milisavljevic}, {Pastorello}, {Sanders}, {Smith}, {Stafford},
  {Thilker}, {Valenti}, {Wood-Vasey}, {Zheng}, {Burgett}, {Chambers},
  {Denneau}, {Draper}, {Flewelling}, {Hodapp}, {Kaiser}, {Kudritzki},
  {Magnier}, {Metcalfe}, {Price}, {Sweeney}, {Wainscoat}, \&
  {Waters}}]{scolnic14}
{Scolnic}, D., {Rest}, A., {Riess}, A., {et~al.} 2014, \apj, 795, 45

\bibitem[{{Sirianni} {et~al.}(2005){Sirianni}, {Jee}, {Ben{\'{\i}}tez},
  {Blakeslee}, {Martel}, {Meurer}, {Clampin}, {De Marchi}, {Ford}, {Gilliland},
  {Hartig}, {Illingworth}, {Mack}, \& {McCann}}]{sirianni05}
{Sirianni}, M., {Jee}, M.~J., {Ben{\'{\i}}tez}, N., {et~al.} 2005, \pasp, 117,
  1049 (S05)

\bibitem[{{Stys} {et~al.}(2004){Stys}, {Bohlin}, \& {Goudfrooij}}]{stys2004}
{Stys}, D.~J., {Bohlin}, R.~C., \& {Goudfrooij}, P. 2004, {Time-Dependent
  Sensitivity of the CCD and MAMA First- Order Modes, Instrument Science
  Report, STIS 2004--04, (Baltimore:STScI)}, Tech. rep.

\bibitem[{{Ubeda} \& {Anderson}(2013)}]{ubeda13}
{Ubeda}, L., \& {Anderson}, J. 2013, {Study of the evolution of the ACS/WFC
  Sensitivity Loss, Instrument Science Report, ACS 2013--01, (Baltimore:
  STScI)}, Tech. rep.

\end{thebibliography}

\begin{deluxetable}{lcccccccccccc}     
\tablewidth{0pt}
\tablecolumns{13}
\tablecaption{Encircled Energy Fractions for Hot Stars at WFC1-1K}
\tablehead{
\colhead{Filter} &\colhead{1 pix} &\colhead{2 pix} &\colhead{3 pix} &\colhead{4 pix}
&\colhead{5 pix} &\colhead{6 pix} &\colhead{7 pix} &\colhead{8 pix} &\colhead{9 pix} 
&\colhead{10 pix} &\colhead{20 pix} &\colhead{40 pix} }
\startdata
 F435W &0.330 &0.663 &0.792 &0.839 &0.863 &0.877 &0.887 &0.895 &0.902 &0.907 &0.941 &0.979\\
 F475W &0.329 &0.670 &0.794 &0.842 &0.868 &0.883 &0.893 &0.901 &0.907 &0.912 &0.944 &0.979\\
 F502N &0.328 &0.670 &0.794 &0.842 &0.869 &0.884 &0.894 &0.902 &0.909 &0.914 &0.945 &0.978\\
 F555W &0.328 &0.668 &0.794 &0.841 &0.868 &0.885 &0.895 &0.903 &0.910 &0.915 &0.946 &0.977\\
 F550M &0.328 &0.666 &0.794 &0.840 &0.867 &0.885 &0.896 &0.904 &0.910 &0.915 &0.947 &0.976\\
 F606W &0.328 &0.661 &0.795 &0.839 &0.866 &0.885 &0.896 &0.904 &0.910 &0.916 &0.947 &0.975\\
 F625W &0.330 &0.655 &0.795 &0.838 &0.864 &0.884 &0.896 &0.904 &0.911 &0.916 &0.948 &0.974\\
 F658N &0.331 &0.651 &0.794 &0.838 &0.863 &0.883 &0.896 &0.904 &0.911 &0.916 &0.948 &0.973\\
 F660N &0.331 &0.650 &0.794 &0.838 &0.863 &0.883 &0.896 &0.904 &0.911 &0.916 &0.948 &0.973\\
 F775W &0.329 &0.625 &0.783 &0.836 &0.858 &0.877 &0.894 &0.904 &0.910 &0.916 &0.949 &0.972\\
 F814W &0.322 &0.611 &0.770 &0.830 &0.853 &0.871 &0.889 &0.901 &0.908 &0.914 &0.949 &0.972\\
 F892N &0.278 &0.546 &0.705 &0.787 &0.818 &0.840 &0.860 &0.877 &0.889 &0.897 &0.942 &0.970\\
F850LP &0.268 &0.532 &0.690 &0.776 &0.810 &0.833 &0.853 &0.871 &0.884 &0.893 &0.940 &0.970\\
\enddata
\label{table:eewfc}
\end{deluxetable}

\begin{deluxetable}{lcccccccccccc}     
\tablewidth{0pt}
\tablecolumns{13}
\tablecaption{Encircled Energy Fractions for Hot Stars at HRC Center}
\tablehead{
\colhead{Filter} &\colhead{2 pix} &\colhead{4 pix} &\colhead{6 pix} &\colhead{8 pix}
&\colhead{10 pix} &\colhead{12 pix} &\colhead{14 pix} &\colhead{16 pix} &\colhead{18 pix} 
&\colhead{20 pix} &\colhead{40 pix} &\colhead{80 pix} }
\startdata
 F220W &0.518 &0.694 &0.755 &0.779 &0.799 &0.818 &0.833 &0.845 &0.857 &0.868 &0.948 &0.977\\
 F250W &0.537 &0.715 &0.783 &0.813 &0.827 &0.842 &0.855 &0.865 &0.874 &0.884 &0.946 &0.980\\
 F330W &0.549 &0.740 &0.804 &0.840 &0.853 &0.864 &0.875 &0.883 &0.891 &0.898 &0.943 &0.981\\
 F344N &0.549 &0.742 &0.806 &0.842 &0.855 &0.866 &0.876 &0.885 &0.892 &0.899 &0.943 &0.981\\
 F435W &0.547 &0.763 &0.817 &0.854 &0.873 &0.883 &0.891 &0.899 &0.905 &0.910 &0.944 &0.981\\
 F475W &0.543 &0.767 &0.819 &0.855 &0.877 &0.888 &0.896 &0.903 &0.909 &0.914 &0.946 &0.980\\
 F502N &0.540 &0.768 &0.820 &0.855 &0.879 &0.891 &0.898 &0.906 &0.912 &0.916 &0.947 &0.980\\
 F555W &0.535 &0.766 &0.821 &0.854 &0.880 &0.893 &0.900 &0.908 &0.914 &0.919 &0.949 &0.980\\
 F550M &0.532 &0.762 &0.822 &0.853 &0.880 &0.894 &0.901 &0.909 &0.915 &0.920 &0.949 &0.980\\
 F606W &0.526 &0.755 &0.822 &0.851 &0.878 &0.894 &0.902 &0.909 &0.915 &0.920 &0.950 &0.979\\
 F625W &0.519 &0.741 &0.821 &0.847 &0.874 &0.891 &0.901 &0.908 &0.914 &0.919 &0.950 &0.978\\
 F658N &0.512 &0.729 &0.819 &0.844 &0.869 &0.888 &0.898 &0.905 &0.911 &0.917 &0.949 &0.977\\
 F660N &0.512 &0.728 &0.818 &0.843 &0.869 &0.887 &0.898 &0.905 &0.911 &0.917 &0.949 &0.977\\
 F775W &0.468 &0.650 &0.783 &0.807 &0.826 &0.847 &0.863 &0.871 &0.878 &0.884 &0.927 &0.964\\
 F814W &0.444 &0.613 &0.756 &0.782 &0.800 &0.821 &0.838 &0.847 &0.855 &0.862 &0.910 &0.956\\
 F892N &0.351 &0.502 &0.642 &0.683 &0.701 &0.720 &0.739 &0.754 &0.764 &0.773 &0.844 &0.925\\
F850LP &0.333 &0.484 &0.619 &0.664 &0.682 &0.700 &0.720 &0.736 &0.746 &0.756 &0.831 &0.919\\
\enddata
\label{table:eehrc}
\end{deluxetable}

\end{document}